\documentclass[12pt]{article}

\usepackage{epsf}

\usepackage{graphicx}%
\usepackage{color}%

\def\refitem#1{\relax}

 \newfont{\cyrfnt}{wncyr9 scaled 1120}

\begin{document}

\begin{center}
{\bfseries Hadron Resonance Gas Model  with Induced Surface Tension}

\vskip 5mm

V. V. Sagun$^{1,2}$, K. A. Bugaev$^{1 *}$,   A. I. Ivanytskyi$^{1}$,  I. P. Yakimenko$^{3}$,
E. G. Nikonov$^{4}$,  A.V. Taranenko$^{5}$,
C. Greiner$^{6}$, 
D. B. Blaschke$^{5, 7, 8}$, 
and G. M. Zinovjev$^{1}$

\vskip 5mm

{\small
$^1${\it
Bogolyubov Institute for Theoretical Physics, Metrologichna str. 14$^B$, Kiev 03680, Ukraine
}
\\
$^2${\it
CENTRA, Instituto Superior T$\acute{e}$cnico, Universidade de Lisboa,
Av. Rovisco Pais 1, 1049-001 Lisboa, Portugal
}
\\
$^{3}${\it
Department of Physics, Chemistry and Biology (IFM), Link\"oping University, SE-58183 Link\"oping, Sweden
}
\\
$^4${\it
Laboratory for Information Technologies, JINR, Joliot-Curie str. 6, 141980 Dubna, Russia
}
\\
$^5${\it National Research Nuclear University ``MEPhI'' (Moscow Engineering Physics
Institute), Kashirskoe Shosse 31, 115409 Moscow, Russia}
\\
$^6${\it
Institute for  Theoretical Physics, Goethe University, Max-von-Laue-Str. 1, 60438 Frankfurt am Main, Germany
}
\\
$^7${\it Institute of Theoretical Physics, University of Wroclaw, pl. M. Borna 9, 50-204 Wroclaw, Poland}
\\
$^8${\it Bogoliubov Laboratory of Theoretical Physics, JINR Dubna, Joliot-Curie str. 6, 141980 Dubna, Russia}
\\
$*$ {\it
E-mail: Bugaev@th.physik.uni-frankfurt.de
}}
\end{center}

\vskip5mm

\centerline{\bf Abstract}
Here we present a physically transparent generalization of the multicomponent Van der Waals equation of state in the grand canonical ensemble. For the one-component case the third and fourth virial coefficients are calculated analytically. It is shown that an adjustment of  a single model parameter allows us  to reproduce  the third and fourth virial coefficients of the gas of hard spheres  with small deviations from their exact values. A thorough comparison of the compressibility factor and speed of sound of the developed model with the one and two component Carnahan-Starling equation of state is made. It is shown that  the model with the induced surface tension is able to reproduce the results of the  Carnahan-Starling equation of state up to the packing fractions 0.2-0.22 at which the usual  Van der Waals equation of state is inapplicable.  At higher packing fractions the developed equation of state is  softer than the gas of hard spheres and, hence,  it breaks causality in the domain where the hadronic description is expected to be inapplicable. Using this equation of state  we develop an entirely new hadron resonance gas model and apply it to a description of the hadron yield ratios measured at  AGS, SPS, RHIC and ALICE  energies of nuclear collisions.  The achieved quality of the fit  per degree of freedom is about 1.08.  We confirm that the strangeness enhancement factor has a peak at low AGS energies, while at and above the highest  SPS energy of collisions the chemical equilibrium of strangeness is observed. We argue that the chemical equilibrium of strangeness, i.e. $\gamma_s \simeq 1$,   observed  above  the center of mass collision energy 4.3 GeV may be related to the hadronization of quark gluon bags which have the Hagedorn mass spectrum, and, hence, it may be a new signal  for  the onset of deconfinement.

\vskip5mm

\section{Introduction.} 

Investigation of the strongly interacting matter equation of state (EoS)  is  the focus of modern nuclear physics and astrophysics \cite{GenRevEOS,Haensel:2007yy}. 
In the low density limit, the system can be considered as a statistical ensemble of composite particles which are characterized by their mass spectrum and their finite size. 
At higher densities, due to the requirement of antisymmetrisation of the wave function of the fermionic constituents, the Pauli blocking effect occurs which can be effectively modeled by the adoption of an  excluded volume. 
At still higher densities the composites become unbound and undergo a dissociation into their constituents
(Mott effect). 
This effect was described first as an explanation of the insulator-to-metal transition \cite{Mott:1974} which occurs in some metal oxides under pressure. 
It has then been taken over to statistical plasma physics and the physics of liquids.
In these systems, the finite size effects in the equation of state for composites are taken into account by a virial expansion which can be subsumed in a very compact form by the excluded volume concept leading, e.g., to the phenomenologically sucessful Carnahan-Starling EoS \cite{CSeos}, see also
\cite{Ebeling:2008mg,EbelingMottBook}.
The proper description of cluster abundances in nuclear matter requires the account for excluded volume 
effects in the nuclear statistical equilibrium (for instance in supernova EoS 
\cite{Lattimer:1991nc,Shen:1998gq})  
which on the more fundamental level are grounded in the Pauli principle leading to the Mott effect for clusters \cite{Typel:2009sy}.
Most recently, the excluded volume concept has proven essential for the discussion of the deconfinement transition in compact stars, where a stiffening of the nuclear matter EoS due to the finite size of nucleons is
a prerequisite for obtaining a novel type of hybrid star EoS allowing for so-called high-mass twin stars, a 
striking effect of quark deconfinement \cite{Benic:2014jia}, potentially observable by precise mass and radius measurements on pulsars. 

The hadron resonance gas model (HRGM)  is not only a low energy density part of  such an  EoS, but it is also an important tool of heavy ion collisions  phenomenology  which allows one to obtain  the 
parameters of chemical freeze-out (CFO)   from experimental data \cite{PBM06,Stachel:2013zma}. 
For more than two decades the HRGM was based on the Van der Waals EoS with one or two hard-core radii (one radius for mesons and another for baryons).  
Although recently  there appeared new variations on the theme of excluded volume models \cite{Typel16} which allow a rather flexible modeling, in particular,  of the nuclear EoS at supranuclear densities, the  main phenomelogical results 
were obtained by 
the HRGM with several hard-core radii  \cite{Horn,KABOliinychenko:12,SFO,Veta14,Bugaev:2014,Bugaev:2015,Bugaev:2016,Bugaev:2016ujp}. Indeed,  such a model  not only provides  the  high quality description  of the hadron multiplicities from the lowest AGS collision energy of a few GeV  to 
the LHC ALICE data measured at the center of mass collision energy $\sqrt{s_{NN}} = 2.76$ TeV,  
but it also allows one to study  the  subtlest features  of hadron matter  thermodynamics at CFO  with very high confidence  \cite{Horn,KABOliinychenko:12,SFO,Veta14,Bugaev:2014,Bugaev:2015,Bugaev:2016,Bugaev:2016ujp}.
Moreover, it  allows one  to find  the novel irregularities and to suggest  the new  signals of the mixed phase formation in the nuclear collisions \cite{Bugaev:2014,Bugaev:2015,Bugaev:2016a}. 

Alas, the presently existing $N$-component HRGM requires to solve $N$ transcendental equations which may include contributions of  hundreds  of corresponding hadronic species. Therefore, further increasing the number of hard-core radii will lead to an essential increase in computational time which does not appear feasible. Moreover, in view of  future experiments at the NICA-JINR and FAIR-GSI accelerators we expect to obtain many more experimental data with, hopefully,  a higher accuracy. These new data will, in principle, allow us to study the second virial coefficients of the most abundant hadrons. 
Therefore, the development of a new multicomponent  HRGM is necessary. 
The validity of this conclusion was demonstrated once more  in \cite{Bugaev:2016} where a thorough analysis of the ALICE data within the conventional HRGM and with the bag-like prescription for hard-core radii suggested in \cite{Vovch15} was performed. 

Another restriction to use the HRGM appears due to the fact that the Van der Waals approximation accounts for the second virial coefficients only and, consequently,  it  can be safely  applied  to low densities, i.e. for packing fractions $\eta = \sum\limits_{all~hadrons} \rho_h  V_h \le 0.11$ \cite{SSpeed} where $V_h$ is the eigenvolume of hadron $h$ and $\rho_h$ is its particle number density. It is also necessary to note  that  the HRGM is the discrete part of the mass-volume spectrum of  quark-gluon bag model with surface tension which has a tricritical \cite{QGBST3crit} or a critical 
\cite{QGBSTcrit} endpoint.  The quark-gluon bag model with surface tension model  allows one to describe the properties of strongly interacting matter at high energy densities. In contrast to the HRGM the continuous part of the QGBSTM  mass-volume spectrum, which describes the large and heavy  quark-gluon-plasma bags employs  the eigenvolume approximation. 
Note that  this  approximation is also used  in all models  describing the  large and heavy  quark-gluon-plasma bags
\cite{QGBST3crit,QGBSTcrit,Carsten07,Carsten08,Ferroni,Carsten10,IvanytskyiNPA,IvanytskyiPRE,UJP2012}.  
It is applicable at high energy densities or  for $\eta > 0.3-0.4$, while at low packing fractions one has to employ the excluded volume modes which reproduces  the low density virial expansion. Therefore, the HRGM with the Van der Waals approximation, i.e. an excluded volume model  (EVM),  is traditionally used for many years. 
However, at the intermediate packing fractions $0.15 < \eta < 0.2$ the EVM  
may easily  become  inapplicable because of the  superluminal values of the speed of sound \cite{SSpeed,Bugaev:2016,Vovch15}.  Hence, 
 an extension of the HRGM beyond the Van der Waals approximation  is also   necessary to model
 the strongly matter EoS near the region of transition between the hadron matter and  quark gluon plasma.

Therefore, in this work we present a completely new version of the HRGM with the multicomponent hard-core repulsion which, by construction,  is the Van der Waals EoS with the induced surface tension (IST EoS hereafter).  This EoS is based on the virial expansion for multicomponent mixture  and, hence,  it naturally switches between the low and high density limits. Comparing it with the Carnahan-Starling EoS \cite{CSeos} for one and two particle species we find almost a perfect coincidence between them up to the packing fractions $\eta \simeq 0.2$-$0.22$. 
Its great advantage is that independently of the number of  different hard-core radii  the IST EoS  is a system of only two transcendental equations.  
Using the IST EoS we successfully  fit the traditional set of the hadron multiplicity ratios \cite{Horn,SFO,Veta14} measured 
at AGS, SPS,  RHIC and ALICE energies of collisions. 

The work is organized as follows. In Section 2 we present the IST EoS, calculate the third and fourth virial coefficients for the one-component case and compare this EoS  with the Carnahan-Starling EoS.  
In Section 3 the necessary formalism is given and the results of fitting the hadron yields ratios are discussed. 
Our conclusions are summarized in Section 4.  A heuristic derivation of  the IST EoS is given in Appendix.


\section{HRGM with the induced surface tension}

A high quality description of  the hadron yield ratios achieved in the last couple of years 
by the  HRGM with multicomponent  hard-core repulsion   is a clear   evidence of its great advantage over the one- and two- component versions.
However, the main disadvantage of  such a  model  is its mathematical complexity, which leads 
to  an essentially longer time of numerical simulations. 
The HRGM \cite{Veta14} is a system of $N$  transcendental equations, where $N$ is the number of  employed  hard-core radii.
Since  in the   HRGM all mesons, except for pions $R_\pi$=0.1 fm and kaons $R_K$=0.395 fm, have single hard-core radius  $R_m \simeq 0.4$ fm  and all baryons, except for 
$\Lambda$-hyperons  $R_{\Lambda}$=0.11 fm, have their own value of hard-core radius $R_{b}$=0.355 fm, then the mesonic and baryonic equations include hundreds of  terms corresponding to accounted hadron states. 
Although these hard-core radii  provide   very high quality description of  111  independent particle yield ratios 
measured in the  central nuclear collisions  for the center of mass collision energies   $\sqrt{s_{NN}} = $ 2.7, 3.3, 3.8, 4.3, 4.9, 6.3, 7.6, 8.8, 9.2, 12, 17, 62.4, 130 and 200 GeV (for the details see  \cite{Horn,SFO,Veta14}) with 
$\chi^2/dof \simeq 0.95$ \cite{Veta14},
any further  increase of  the number of  independent  hadronic hard-core radii in the HRGM  will face 
severe computational difficulties.
 More details on  the HRGM equations and the list of experimental data   employed  in the  actual simulations   can be found in  \cite{Horn,KABOliinychenko:12,SFO,Veta14}.

Here we employ a more effective model with multicomponent hard-core repulsion named the IST EoS
because besides the hard-core repulsion  it explicitly contains the surface tension induced by the inter particle interaction \cite{Bugaev:13NPA}.
Recently, such an EoS  was 
proposed on the basis of the virial expansion of  multicomponent mixture \cite{Bugaev:13NPA} obtained for the simplified  statistical multifragmentation model  \cite{Mekjian} with an infinite number of hard-core radii of nuclear fragments of all sizes. 
Such an EoS is a system of coupled  equations between the system pressure  $p$ and  the induced surface tension coefficient $\Sigma$ which has the form  \cite{Bugaev:2016,Bugaev:13NPA}  
(for convenience, its simplified derivation is given in Appendix)
\begin{eqnarray}
\label{EqI}
p &=& T \sum_{k=1}^N \phi_k \exp \left[ \frac{\mu_k}{T} - \frac{4}{3}\pi R_k^3 \frac{p}{T} - 4\pi R_k^2 \frac{\Sigma}{T} \right]
\,, \\
\label{EqII}
\Sigma &=& T \sum_{k=1}^N R_k \phi_k \exp \left[ \frac{\mu_k}{T} - \frac{4}{3}\pi R_k^3 \frac{p}{T} - 4\pi R_k^2 \alpha \frac{\Sigma}{T} \right] \,,\\
\label{EqIII}
\mu_k &=& \mu_B B_k + \mu_{I3} I_{3k} + \mu_S S_k \,.
\end{eqnarray}
Here $\mu_B$, $\mu_S$, $\mu_{I3}$ are, respectively, the baryonic,  the strange and the third projection of isospin  chemical potentials, while 
$B_k$, $S_k$, $I_{3k}$, $m_k$ and $R_k$ denote, respectively,  the corresponding charges,  mass and hard-core radius of the $k$-sort of hadrons. The summations in  Eqs. (\ref{EqI}) and (\ref{EqII})  are made over all sorts of hadrons and their antiparticles are 
considered as independent species. 

In Eq. (\ref{EqII}) the dimensionless  parameter $\alpha>1$ is introduced   due to the freedom of the Van der Waals extrapolation to high densities \cite{Bugaev:13NPA}.  As it is shown below,  the  parameter $\alpha$ accounts for the  high density terms of virial expansion  and it allows us to   modify  the  Van der Waals  EoS  to a more realistic one.
In principle, $\alpha$ can be a regular function of $T$ and $\mu$,   however, for the sake of simplicity it is fixed to a constant value. 

The one-particle thermal density $\phi_k$ in Eqs. (\ref{EqI}) and (\ref{EqII})  accounts for  the  Breit-Wigner  mass attenuation and is written in the Boltzmann approximation
\begin{eqnarray}
\label{EqIV}
\phi_k = g_k  \gamma_S^{|s_k|} \int\limits_{M_k^{Th}}^\infty  \,  \frac{ d m}{N_k (M_k^{Th})} 
\frac{\Gamma_{k}}{(m-m_{k})^{2}+\Gamma^{2}_{k}/4} 
\int \frac{d^3 p}{ (2 \pi)^3 }   \exp \left[ -\frac{ \sqrt{p^2 + m^2} }{T} \right] \,,
\end{eqnarray}
for all hadrons except pions. 
Here $g_k$ is the  degeneracy factor of the $k$-sort  of hadrons,
$\gamma_S$ is the strangeness suppression factor \cite{Rafelski}, $|s_k|$ is the number of valence  strange quarks and antiquarks  in this kind of hadrons, 
${N_k (M_k^{Th})} \equiv \int\limits_{M_k^{Th}}^\infty \frac{d m \, \Gamma_{k}}{(m-m_{k})^{2}+\Gamma^{2}_{k}/4} $ denotes 
a corresponding normalization, while $M_k^{Th}$ corresponds to the decay threshold mass of the $k$-sort of hadrons.
We would like to point out the fact that  usage of the   mass attenuation like  in Eq.  (\ref{EqIV})
for a mixture of  hadron resonances
can be rigorously derived   \cite{David:16A,David:16B} from  the Phi-functional approach \cite{Phi-approach},  when the
Phi-functional is chosen from the class of  two-loop diagrams only, see also \cite{Phi-approach2}.

For the pions of sort $A = \{-, 0, + \}$, instead of  the Boltzmann distribution (\ref{EqIV}) we use the Bose-Einstein distribution function 
\begin{eqnarray}
\label{EqV}
\phi_{\pi_A} =  \int \frac{d^3 p}{ (2 \pi)^3 }   \frac{1}{\exp \left[ \frac{ \sqrt{p^2 + m_\pi^2} - \mu_{I3} I_{3,\, A}}{T} \right]  - 1} \,,
\end{eqnarray}
because at high temperatures, which will be analyzed here, the quantum correction cannot be ignored.  Here the particle density of pions depends on the 
 third  projection of  isospin $I_{3,\, A}$ and  the corresponding chemical potential $\mu_{I3}$.

The system of Eqs.  (\ref{EqI}), (\ref{EqII}) and  (\ref{EqIII}) should be supplemented by the strange charge conservation law,
which in case of relativistic collisions of heavy ions has the  form
\begin{eqnarray}
\label{EqVI}
\sum_{k=1}^N \phi_k S_k \exp \left[ \frac{\mu_k}{T} - \frac{4}{3}\pi R_k^3 \frac{p}{T} - 4\pi R_k^2 \frac{\Sigma}{T} \right] = 0 \,,
\end{eqnarray}
which completes the  system of equations employed for analysis of hadron multiplicities.

In the work  \cite{Bugaev:13NPA} it was established that the parameter  $\alpha$ should be  greater than one in order 
to reproduce the physically correct phase diagram properties of nuclear matter.  In order to determine the correct value of parameter   
$\alpha$, we compare here the IST EoS for the  point-like 
pions, for the nucleons and $\Delta (1232)$  baryons having the same hard-core radius of 0.4 fm with the famous Carnahan-Starling EoS  \cite{CSeos} found for the same temperature, same particle densities  and same   hard-core radii.   For simplicity  the antibaryons are neglected in this treatment.

\begin{figure}[htbp]
\centerline{\hspace*{-4.4mm}
\includegraphics[width=104mm]{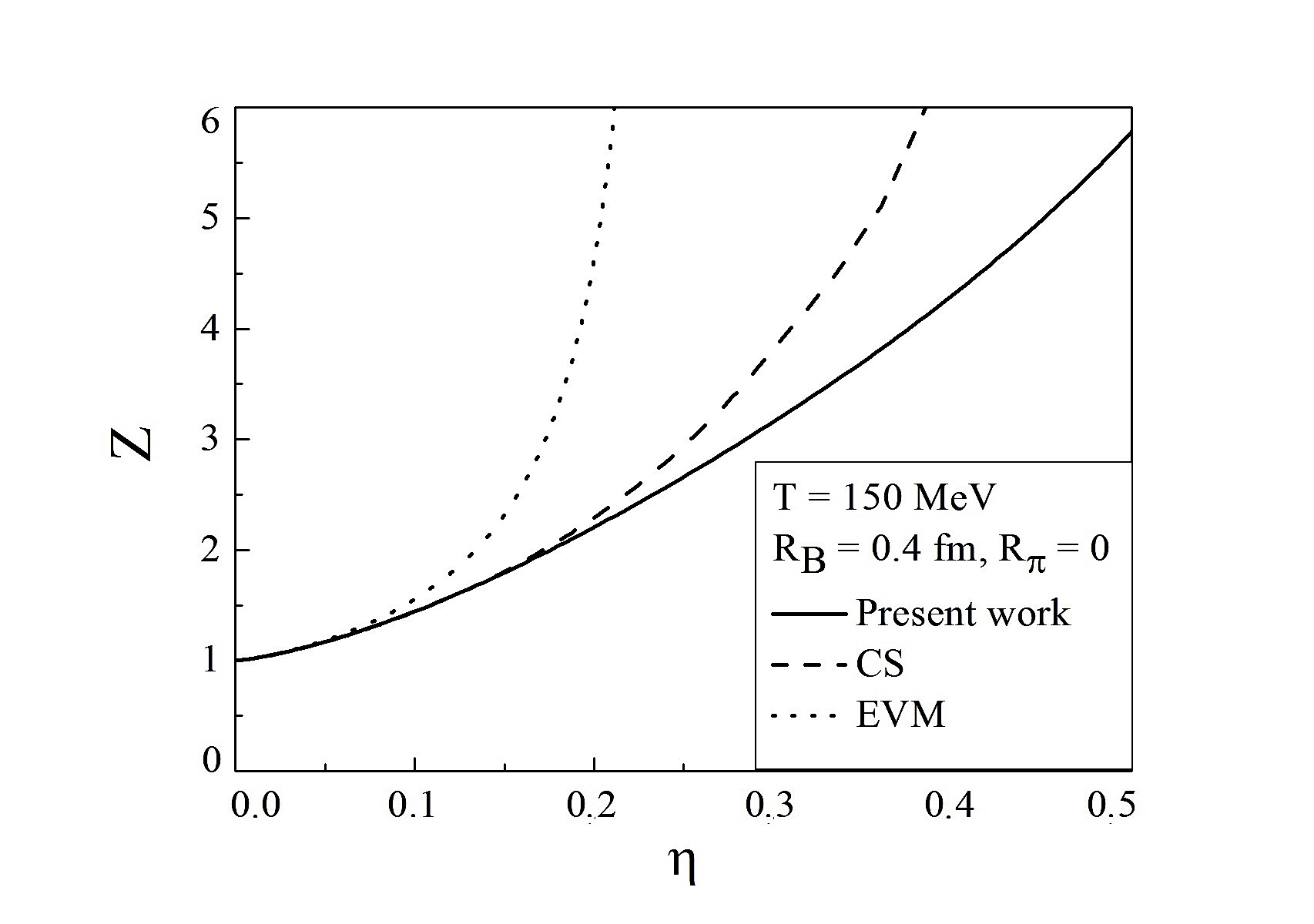}
  \hspace*{-11mm}
\includegraphics[width=92mm]{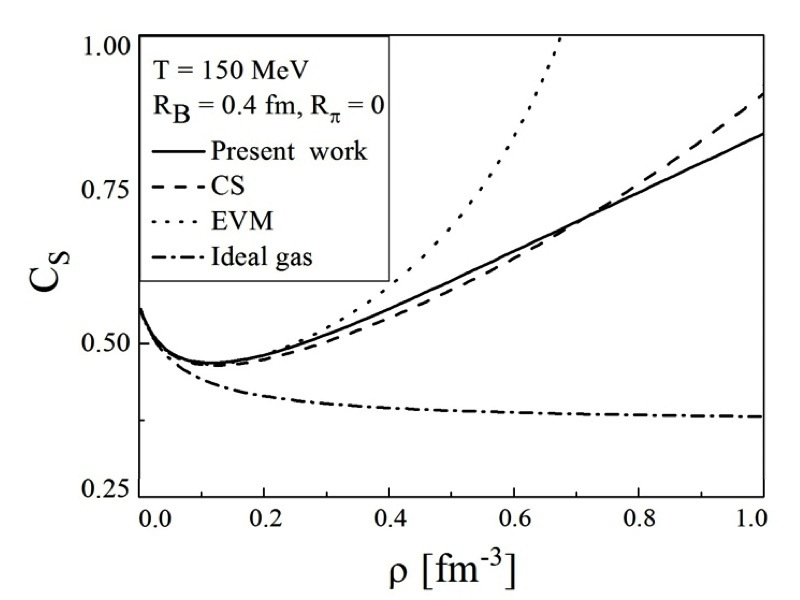}
 } 
 \caption{{\bf Left panel:} Compressibility factor $Z$  of  the gas consisting from the point-like pions,  the nucleons and $\Delta (1232)$  baryons  having the hard-core radius of $0.4$ fm  is shown for different EoS  as a function of baryon packing fraction $\eta$.  The  Van der Waals  EoS (dotted curve), the IST EoS (solid curve)  and CS EoS (long dashed curve) are shown 
  for $T=$ 150 MeV. {\bf Right panel:} The speed of sound as a function of baryonic density is shown for the same EoS as in the left panel  and with the same notations. The dotted-dashed curve shows the   speed of sound for  point-like pions and baryons.}
  \label{Fig1}
\end{figure}

\begin{figure}[htbp]
\centerline{\hspace*{-4.4mm}
\includegraphics[width=104mm]{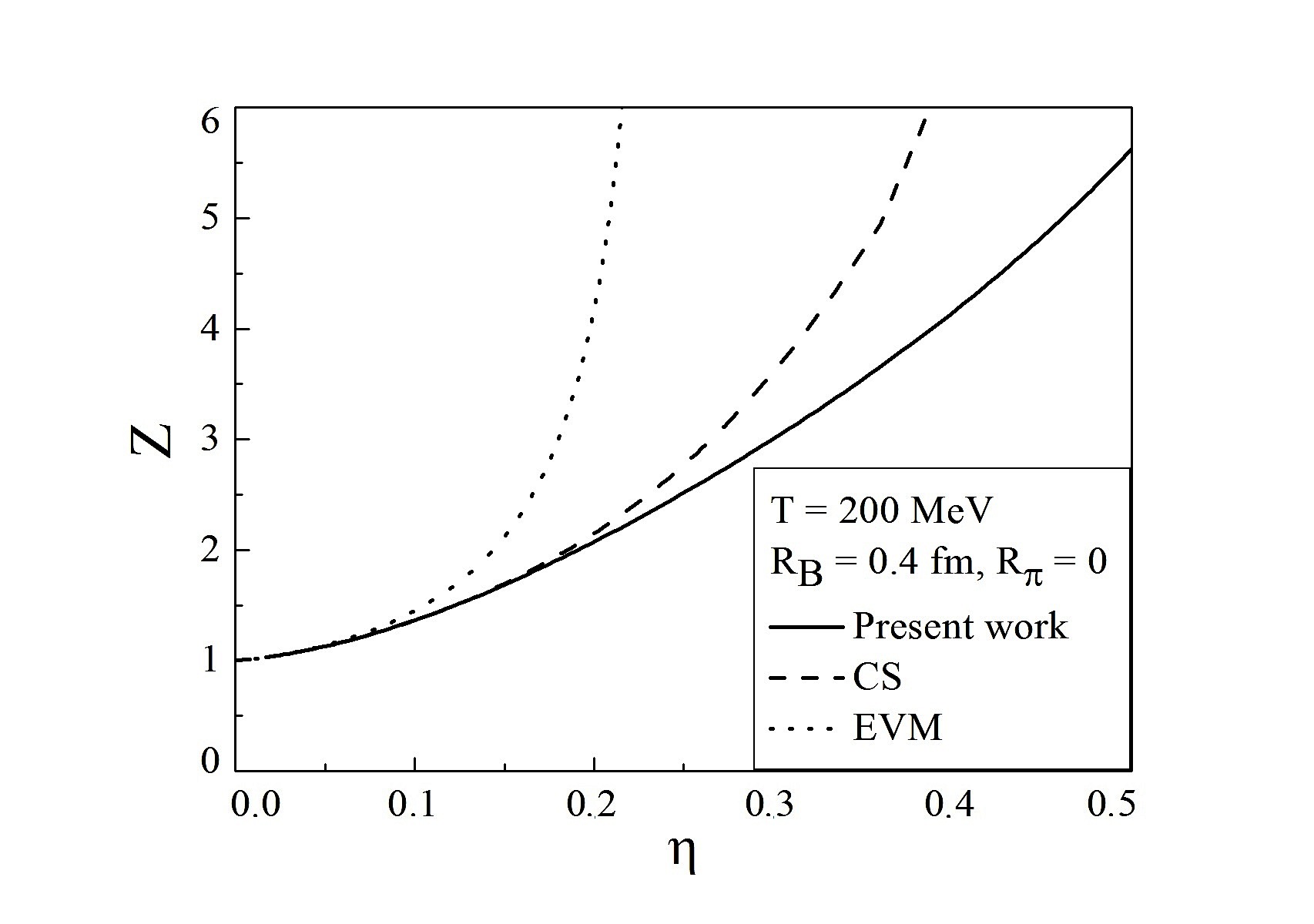}
  \hspace*{-11mm}
\includegraphics[width=92mm]{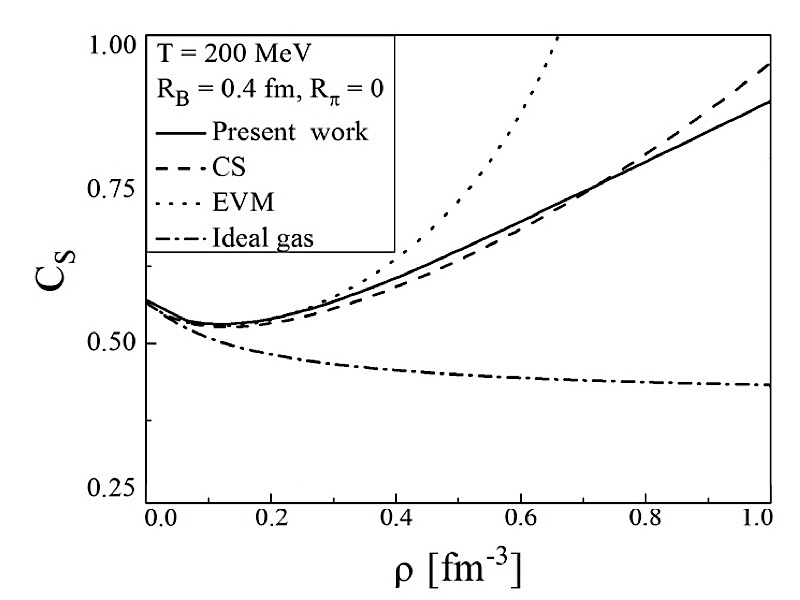}
 } 
 \caption{Same as in Fig. \ref{Fig1}, but for  $T=$ 200 MeV. }
  \label{Fig2}
\end{figure}

In order to further  simplify a comparison we consider the case  $\mu_{I3} = \mu_S =0$, while $\mu_B > 0$.  Then the 
 Carnahan-Starling  EoS \cite{CSeos} for the mixture of baryons (the nucleons and $\Delta (1232)$  baryons) and point-like pions is   \cite{SSpeed}
\begin{eqnarray}
\label{EqVII}
P^{CS}~=~{ 3 T \phi_{\pi_0} (T)}  +  \rho_B\, T  \, Z_B, 
\hspace*{0.2cm} Z_B = \frac{1+\eta +\eta^2 - \eta^3}{(1 - \eta)^{3}} \,,
\end{eqnarray}
where the packing fraction of baryons  of the same hard-core radius $R$  is $\eta =  \frac{4}{3} \pi R^3 \, \rho_B$ and  $\rho_B$ is   their  baryon density $\rho_B \equiv \frac{\partial p}{\partial \mu_B}$ and pion number density $\rho_\pi  \equiv  \frac{\partial  p}{\partial \mu_\pi} \biggl|_{\mu_\pi = 0}$ can be  found from  Eq. (\ref{EqVII}) and from the system (\ref{EqI})-(\ref{EqV})  for a comparison.  
More details on calculating the particle densities and speed of sound  for Eq.  (\ref{EqVII}) can be found  in  \cite{SSpeed} (see Eqs. (59)-(64)), while the necessary expressions for the particle densities of the IST EoS are given below  (see Eqs. (\ref{EqXXI}) and (\ref{EqXXVI})).
As one can see from Figs. \ref{Fig1} and 
\ref{Fig2}    the IST EoS  with $\alpha=$1.25 reproduces the compressibility factor 
$Z \equiv \frac{p}{T (\rho_B + 3 \rho_{\pi_0} (T))}$  up to  
$\eta \simeq$ 0.22.  From these figures one can  also see that the speed of sound $c_S$ of  (\ref{EqVII}) is almost perfectly reproduced by the IST EoS with $\alpha=$1.25 up to five values of normal nuclear density ($\eta \simeq 0.22$).  These figures also show 
that the EVM can be used up to $\eta \simeq$ 0.11 \cite{SSpeed}. 

Note that such a comparison is justified by the fact that the hard-core radii found by the HRGM in \cite{Veta14} are close to 0.4 fm, except for pions and $\Lambda$-hyperons which are about 0.1 fm, i.e. their eigenvolume   is about  $4^3 = 64$ times smaller than the one of other hadrons. 
Although  these hard-core radii were found within the range of applicability of the EVM  approximation, we expect that  fitting the same data with the IST EoS (\ref{EqI})-(\ref{EqV})  will not drastically change their values. 

In order to reveal the reason for such a good correspondence between the Carnahan-Starling EoS and the IST EoS with 
$\alpha=$1.25 we   calculate the third and fourth virial coefficients of the system  (\ref{EqI})-(\ref{EqV}) for the same hard-core radius $R$ and for the same (baryonic) charge of particles, i.e. for  $\{R_k\} =R$  and $\{B_k\} =1$ for any $k$.  Then differentiating Eqs. (\ref{EqI}) and (\ref{EqII}) with respect to 
$\mu_k = \mu$ and excluding from them the derivative
\begin{eqnarray}\label{EqVIII}
\frac{\partial \Sigma}{\partial \mu} &=&  \frac{\Sigma}{T} \,  \frac{(1 - v \rho)}{1 + \frac{\alpha\, s\, \Sigma}{T}} \,,
\end{eqnarray}
one can find the following expression for particle density 
\begin{eqnarray}\label{EqIX}
\rho & \equiv & \frac{\partial p}{\partial \mu} ~=~  \frac{p}{T} \, (1 - v \rho)  \, 
\frac{1 + \frac{(\alpha-1)\, s\, \Sigma}{T}}{1 + \frac{\alpha\, s\, \Sigma}{T}}
 \,.
\end{eqnarray}
Here $v = \frac{4}{3}\pi R^3$ and $s = 4\pi R^2 $ are, respectively, the eigenvolume and eigensurface of particles of hard-core radius $R$.  Dividing Eq. (\ref{EqII}) for $\{R_k=R\}$ by Eq. (\ref{EqI}) for $\{R_k=R\}$, one can establish  a useful identity
\begin{eqnarray}
\label{EqX}
\Sigma  & =   &  p \, R  \, \exp \left[ - s \cdot (\alpha-1)\, \frac{\Sigma}{T} \right] \equiv p \, R \, E_\Sigma \, . 
\end{eqnarray}
Using  the identity  (\ref{EqX}), one can identically rewrite  (\ref{EqIX}) in the form 
\begin{eqnarray}\label{EqXI}
p & = &   \frac{T \,\rho}{\left[  1 - v \rho -  \frac{3\, v\, \rho \, E_\Sigma }{1 + \frac{\left(\alpha-1 \right)\, 3\, v \, p E_\Sigma}{T} } 
\right]} \\
\label{EqXII}
&\equiv  &  \frac{T \,\rho}{\left[  1 - v^{eff} \rho \right]} \, .
\end{eqnarray}
Expression  (\ref{EqXI})  is  convenient for further evaluation.  Here we used an evident relation $s \, R = 3 \, v$.  From the denominator of  (\ref{EqXI})  one can find an effective excluded volume of particle
\begin{eqnarray}
\label{EqXIII}
v^{eff}   & =   & v  \left[1 +   \frac{3\,  E_\Sigma }{1 + \frac{\left(\alpha-1 \right)\, 3\, v \, p E_\Sigma}{T}}  \right] \,.
\end{eqnarray}
This expression shows that at low densities, i.e. for  $\frac{|\alpha-1| \, s\, \Sigma }{T} \ll 1$, and from (\ref{EqX})  one obtains  $E_\Sigma \rightarrow 1$; and hence  $v^{eff} \simeq 4 \,v$   correctly reproduces the excluded volume.  
In the  high density limit, on the other hand, $\frac{p\, v }{T}  \gg 1$ and, hence,
$\frac{(\alpha-1) \, s\, \Sigma }{T}  \gg 1$  since $\mu/T \rightarrow\infty$. Therefore, for $\alpha >1$  one finds that $E_\Sigma \rightarrow 0$ and the effective excluded volume in  Eq. (\ref{EqXII})  becomes equal to the eigenvolume, i.e. $v^{eff} \simeq  v$. Thus, in the IST EoS  the parameter  $\alpha$ switches  between the excluded volume and the eigenvolume regimes.
Also from  Eqs.  (\ref{EqX})  and  (\ref{EqXII})  one can find that for $\alpha =1$ it follows that  $v^{eff} =  4 \,v$, i.e. the IST EoS recovers the usual EVM result. 

Expanding the denominator in (\ref{EqXI})  in a geometric series, expanding $E_\Sigma$  in the Taylor series in powers of $\Sigma$, and applying the aforementioned identity (\ref{EqX})  to them one can get the following result for the pressure of the one-component Boltzmann gas 
\begin{eqnarray}
\label{EqXIV}
p & \simeq   & T \rho ( 1 + 4 \, v \rho + B_3 \,\rho^2 + B_4 \, \rho^3 + ...) \,, \\
\label{EqXV}
B_3 & = &  \left[16 - 18 (\alpha-1) \right] v^2 \,, \\
 \label{EqXVI}
B_4 & = &  \left[64 + \frac{243}{2}(\alpha-1)^2 - 216 (\alpha-1) \right] v^3 \,,
\end{eqnarray}
where in the intermediate steps we substituted the virial expansion  (\ref{EqXIV}) into the right hand side of  (\ref{EqXI}). 
Comparing this result with the virial expansion of the one-component gas of  hard spheres \cite{BookLiquids}
\begin{eqnarray}
\label{EqXVII}
p & \simeq   & T \rho ( 1 + 4 \, v \rho + 10 \,v^2 \rho^2 + 18.36 \,  v^3  \rho^3 + ...) \,, 
\end{eqnarray}
one finds that $B_3 (\alpha =4/3) = 10\, v^2$, but, unfortunately, in this case   $B_4 (\alpha =4/3) \simeq 5.5\, v^3$, which is too small compared to $18.36 \,  v^3$. 
On the other hand, solving an equation 
\begin{eqnarray}\label{EqXVIII}
B_4 (\alpha) = 18.36 \,  v^3 \,, 
\end{eqnarray}
 one finds two solutions $\alpha_1 \simeq 1.245$  and
$\alpha_2 \simeq 2.533$. Since  $B_3 (\alpha =\alpha_1) \simeq 11.59\, v^2$ and $B_3 (\alpha =\alpha_2) \simeq - 11.59\, v^2$,
it is evident that  the correct root is $\alpha=\alpha_1 \simeq 1.245$. 
This is an indication of a very good correspondence between the Carnahan-Starling EoS  (\ref{EqVII}) and the IST EoS with $\alpha=1.25$; in fact, $B_3 (\alpha =1.25) \simeq 11.5\, v^2$ and $B_4 (\alpha =1.25) \simeq 17.59\, v^3$. In other words,  the one-component IST EoS reproduces the third virial coefficient of the gas of hard spheres with the relative error $+15$\%  and the fourth virial coefficient with  the relative error  $-4.1$\%.  Note that  at the packing fraction $\eta = v \rho =0.2$ the deviation of the compressibility factor $Z(\eta) = \frac{p}{T \, \rho}$ generated by these errors is, respectively,  $1.5 \eta^2 \simeq 0.06$ and $-0.766  \eta^3 \simeq -0.006$, which should be compared to the value $Z(0.2) \simeq 2$. In other words,  at the packing fraction $\eta =0.2$ the relative deviation  of the IST EoS $Z(0.2)$ from the one of hard spheres  is less than 3\%! Clearly, for $\eta < 0.2$ the deviation of  the IST EoS from the EoS of hard spheres and from the Carnahan-Starling EoS  is  much smaller.  

\begin{figure}[htbp]
\centerline{ \hspace*{11mm}
\includegraphics[width=92mm,height=68mm]{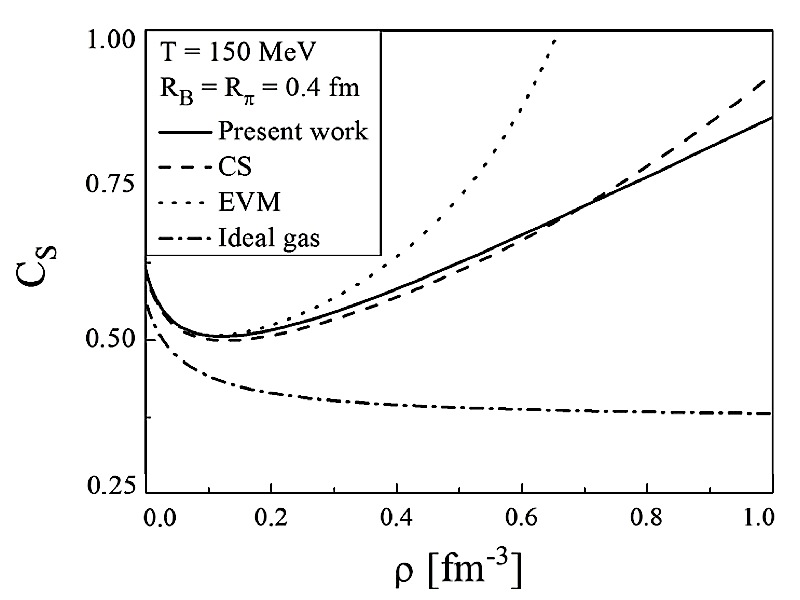}
  \hspace*{-4mm}
\includegraphics[width=105mm]{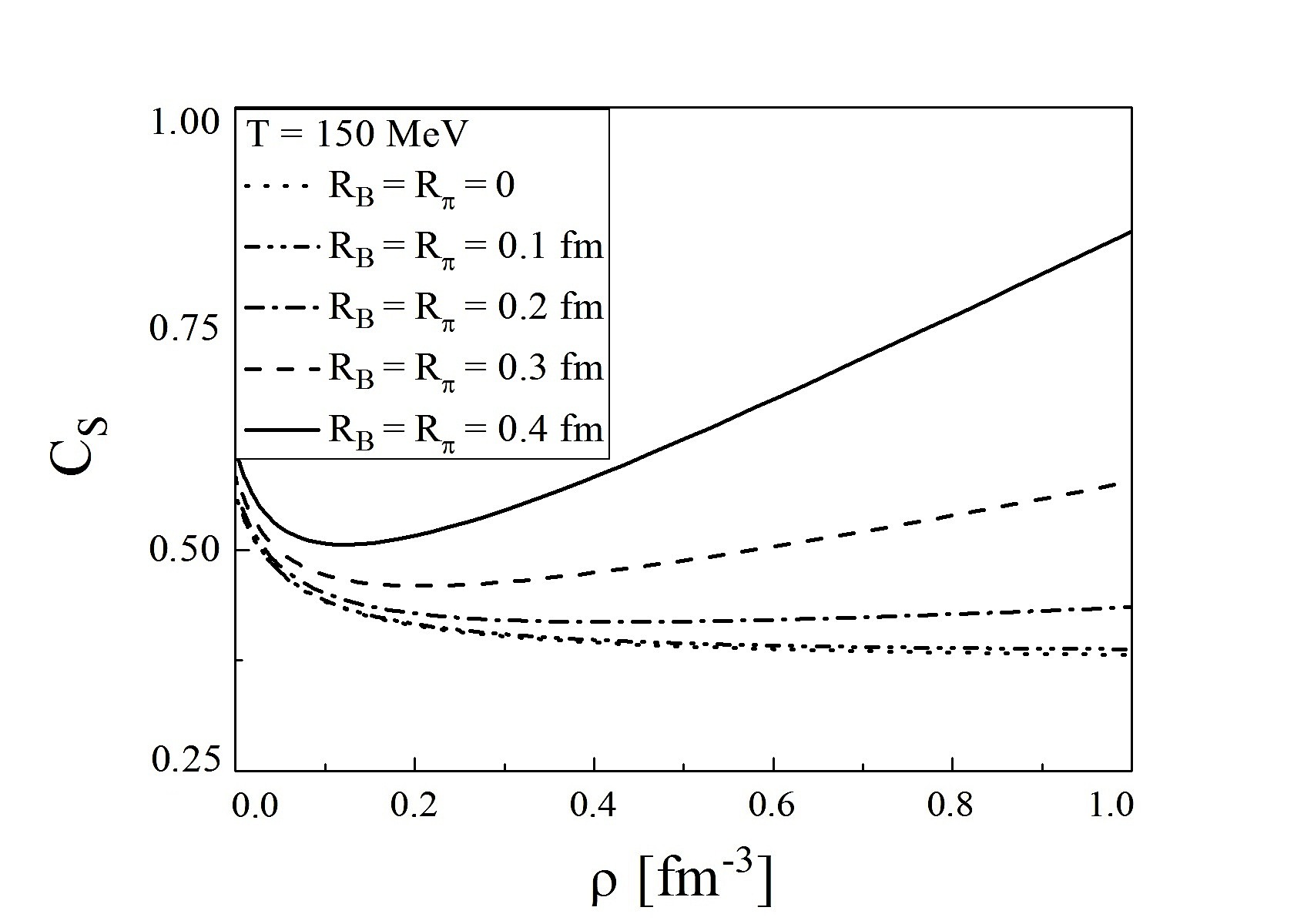}
 } 
 \caption{{\bf Left panel:}The speed of sound of  the gas consisting from the pions,  the nucleons and $\Delta (1232)$  baryons  having the same hard-core radius of $0.4$ fm  is shown for different EoS.  The  Van der Waals  EoS (dotted curve), the IST EoS (solid curve)  and CS EoS (long dashed curve) are shown 
  for $T=$ 150 MeV. {\bf Right panel:} The speed of sound of one-component IST EoS  as a function of baryonic density is shown for the same gas as in the left panel, but  for   different values of hard-core radius.}
  \label{Fig3}
\end{figure}

To illustrate the validity of this conclusion in the left panel of  Fig. \ref{Fig3}
we compare the Carnahan-Starling EoS and the IST EoS with $\alpha =1.25$ for  the nucleons, $\Delta (1232)$  baryons  and pions of the same hard-core radius $R=0.4$ fm. As one can see from this figure the coincidence of these two EoS is the same as for the Carnahan-Starling EoS with point-like pions demonstrated in Figs.  \ref{Fig1} and \ref{Fig2}.
Thus,  here we showed that the IST EoS with a single additional parameter $\alpha$ compared to the EVM is not only able to reproduce the second virial coefficient, but it is also able  to reproduce the third and fourth  virial coefficients of the gas of  hard spheres very well! 
Therefore, in this work we employ the value  $\alpha =1.25$.  

From the right panel of   Fig. \ref{Fig3} one can see that  the one-component IST  EoS gets softer and the region of  its causality widens, if  the common hard-core radius of hadrons  decreases.

\section{Fitting the hadron yield ratios}

The fit procedure along with the detailed description of  the experimental set of 111 independent ratios  used here is very well documented in  \cite{Horn,Veta14,Bugaev:2016ujp}  and, hence, there is no reason to repeat this well known information. 
The data are the ratios measured in the  central nuclear collisions  for the center of mass collision energies    $\sqrt{s_{NN}} = $ 2.7, 3.3, 3.8, 4.3, 4.9 (AGS energies), 6.3, 7.6, 8.8, 12, 17 (NA49 data), 9.2, 62.4, 130 and 200 GeV (STAR data).  Here we do not analyze  the Beam  Energy Scan data measured at RHIC, since these data are rather preliminary and, consequently, they often have rather large error bars even for the hadronic multiplicities, hence,  we do not think that a fit of these data will give us any new information about the thermodynamics of CFO.   Nevertheless,  in addition to the data measured at AGS, SPS and RHIC energies, we analyze the ALICE data  \cite{Abelev:2013vea,Abelev:2013zaa,Abelev:2013xaa,Knospe:2013tda,Adam:2015vda,Donigus:2015bsa,Adam:2015yta}
from which the independent ratios were prepared in  \cite{Bugaev:2016}. 

A few remarks should be made  here about the ALICE data used in a fit.
Similarly to Ref.  \cite{Stachel:2013zma} we 
do not include the $K^*$ data into the fitting procedure,  since the reactions like $K+\pi \leftrightarrow K^*$ can occur after chemical freeze-out and change the $K^*$ yields \cite{Stachel:2013zma}. For  a detailed explanation see the caption of Fig. 1 in \cite{Stachel:2013zma}. 
However, in contrast to \cite{Stachel:2013zma} we do not include into a fit the ratios which  involve  the (anti)nuclei. Although, similarly to  \cite{Stachel:2013zma}, it is possible to fit the full set of the ALICE data \cite{Abelev:2013vea,Abelev:2013zaa,Abelev:2013xaa,Knospe:2013tda,Adam:2015vda,Donigus:2015bsa,Adam:2015yta} with the HRGM \cite{Bugaev:2016},
we do not  think that taking the hard-core radius of  (anti)nuclei to be the same as for baryons is a correct approach.

\begin{table}
\begin{center}
  \begin{tabular}{ccc}
          Ratio                                              &   Value         & Error     \\
          \hline \\
          $\pi^-/\pi^+$                                      &        0.99776  &   0.10023 \\
          $K^-/K^+$                                          &        0.99500  &   0.11645 \\
          $\bar{p}/p$                                        &        0.98387  &   0.11313 \\
          $\Xi^-/\Xi^+$                                      &        1.01829  &   0.10552 \\
          $\Omega^-/\Omega^+$                                &        0.96667  &   0.23371 \\
          $\phi/K^-$                                         &        0.11250  &   0.02500 \\
          $p/\pi^+$                                          &        0.04630  &   0.00500 \\
          $K^+/\pi^+$                                        &        0.14937  &   0.01605 \\
          $\Lambda/\pi^+$                                    &        0.03585  &   0.00453 \\
          $\Xi^+/\pi^+$                                      &        0.00490  &   0.00050 \\
          $\Omega^+/\pi^+$                                   &        0.00090  &   0.00016 \\

  \end{tabular}
  \caption{Ratios which were obtained in \cite{Bugaev:2016} and which   are  analyzed here.}
  \label{tab:rat_fit}
  \end{center}
\end{table}

\subsection{Necessary Formalism}

To fit the ratios we need the explicit expressions for the particle number densities. 
Introducing the partial pressure $p_k$ and the partial surface tension coefficient $\Sigma_k$ of a hadron of sort  $k$
\begin{eqnarray}
\label{EqXIX}
p_k &=& T \phi_k \exp \left[ \frac{\mu_k}{T} - \frac{4}{3}\pi R_k^3 \frac{p}{T} - 4\pi R_k^2 \frac{\Sigma}{T} \right]
\,, \\
\label{EqXX}
\Sigma_k &=& T  R_k \phi_k \exp \left[ \frac{\mu_k}{T} - \frac{4}{3}\pi R_k^3 \frac{p}{T} - 4\pi R_k^2 \alpha \frac{\Sigma}{T} \right]  \equiv  p_k R_k  \exp\left[ - 4\pi R_k^2  (\alpha -1) \frac{\Sigma}{T}   \right] \,,
\end{eqnarray}
one can get the  particle number  density of hadrons of sort $k$ as
\begin{equation}\label{EqXXI}
\rho_k \equiv \frac{\partial  p}{\partial \mu_k} = \frac{1}{T} \cdot \frac{p_k \, a_{22} - \Sigma_k \, a_{12}}{a_{11}\, a_{22} - a_{12}\, a_{21} } \,,
\end{equation}
where the coefficients $a_{kl}$ can be expressed in terms of  the partial pressures $\{ p_k\}$ and the partial surface tension coefficients $\{\Sigma_k\}$ as
\begin{eqnarray}
a_{11} &=& 1 + \frac{4}{3} \pi \sum_k  R_k^3 \frac{p_k}{T} \,, \\
a_{12} &=& 4 \pi \sum_k R_k^2 \frac{p_k}{T} \, , \\
a_{21} &=& \frac{4}{3} \pi \sum_k R_k^3\frac{\Sigma_k}{T} \, ,\\
a_{22} &=& 1 + 4 \pi  \alpha \sum_k  R_k^2 \, \frac{\Sigma_k}{T} \,.
\end{eqnarray}
In case of  the pion-like particles, say pions ($R_\pi = 0)$,  the expression  (\ref{EqXXI}) is simplified as 
\begin{eqnarray}\label{EqXXVI}
\rho_\pi & \equiv & \frac{\partial  p}{\partial \mu_\pi} \Biggl|_{\mu_\pi = 0} =  \frac{1}{T} \cdot \frac{p_\pi }{a_{11} - \frac{a_{12}\, a_{21}}{a_{22}} }  \,, \\
\rho_\pi  & \underbrace{\longrightarrow}_{low~densities} &  \frac{1}{T} \cdot \frac{p_\pi }{1 +  \frac{4}{3} \pi \sum_k  R_k^3 \frac{p_k}{T}}  \,,
\label{EqXXVII}
\end{eqnarray}
which can be  simplified further at low densities, as it is seen  from Eq. (\ref{EqXXVII}), since in this limit  one can  safely 
neglect  in (\ref{EqXXVI})   the product $a_{12}\, a_{21}$  compared to $a_{22}$.
This expression demonstrates the meaning of the hard-core interaction for point-like particles. Indeed, at low densities   the partial  pressure of  each hadron  is close to the ideal gas one, i.e. $p_k \simeq T \rho_k$,  and, hence,  a sum in  the  denominator of  Eq. (\ref{EqXXVII}) accounts  for the fact that the point-like pions cannot occupy the volume which is already filled up by the eigenvolumes of all other particles.  We note that Eqs. (\ref{EqXXI}) and  (\ref{EqXXVI}) were used, respectively,   to evaluate the density of baryons and  point-like pions for a comparison  between the IST EoS and  the EoS given by  (\ref{EqVII}).

\begin{figure}[htbp]
  \centerline{\hspace*{7.7mm}
    \includegraphics[height=70mm]{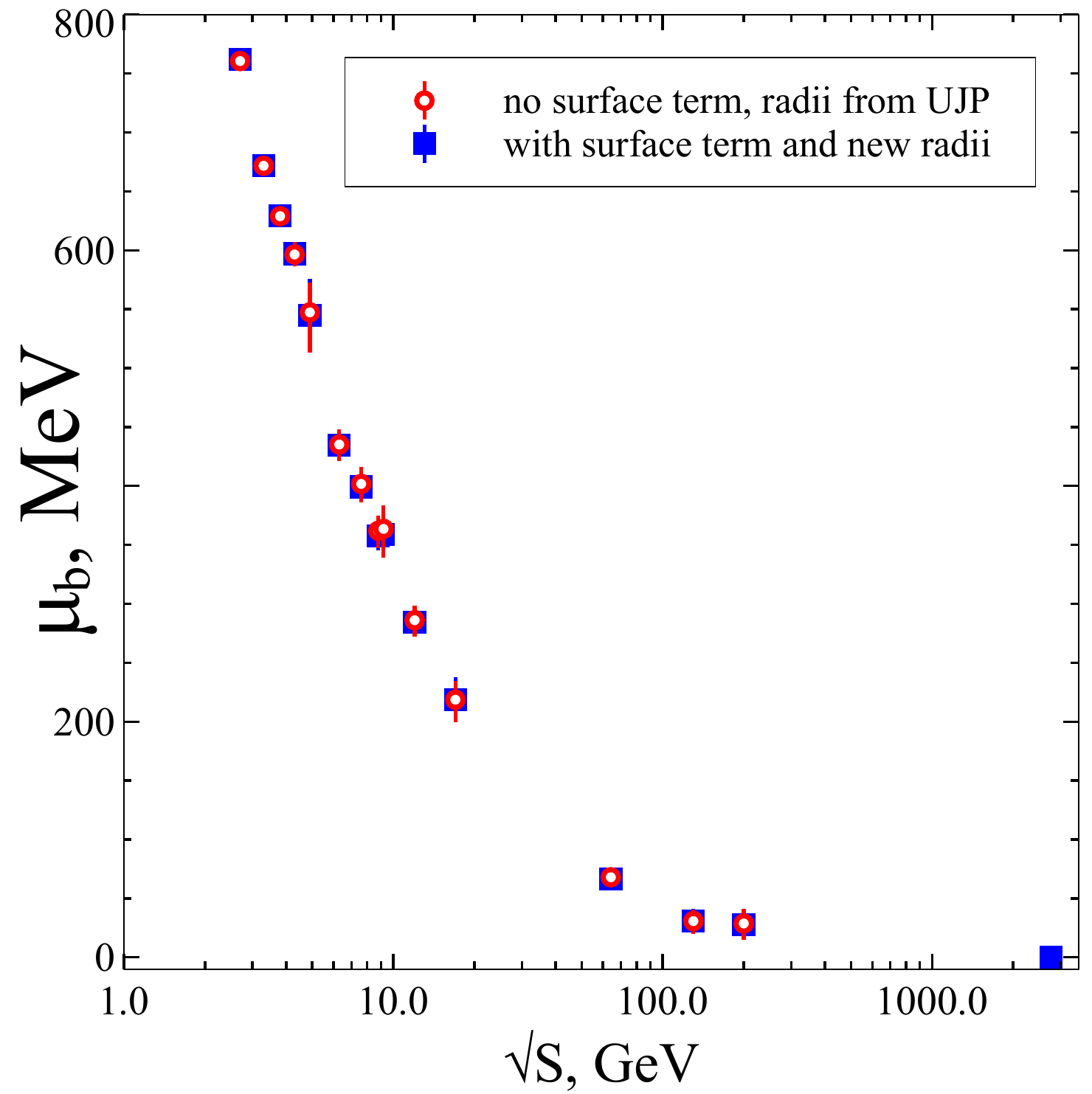}
  \hspace*{11mm}
   \includegraphics[height=70mm]{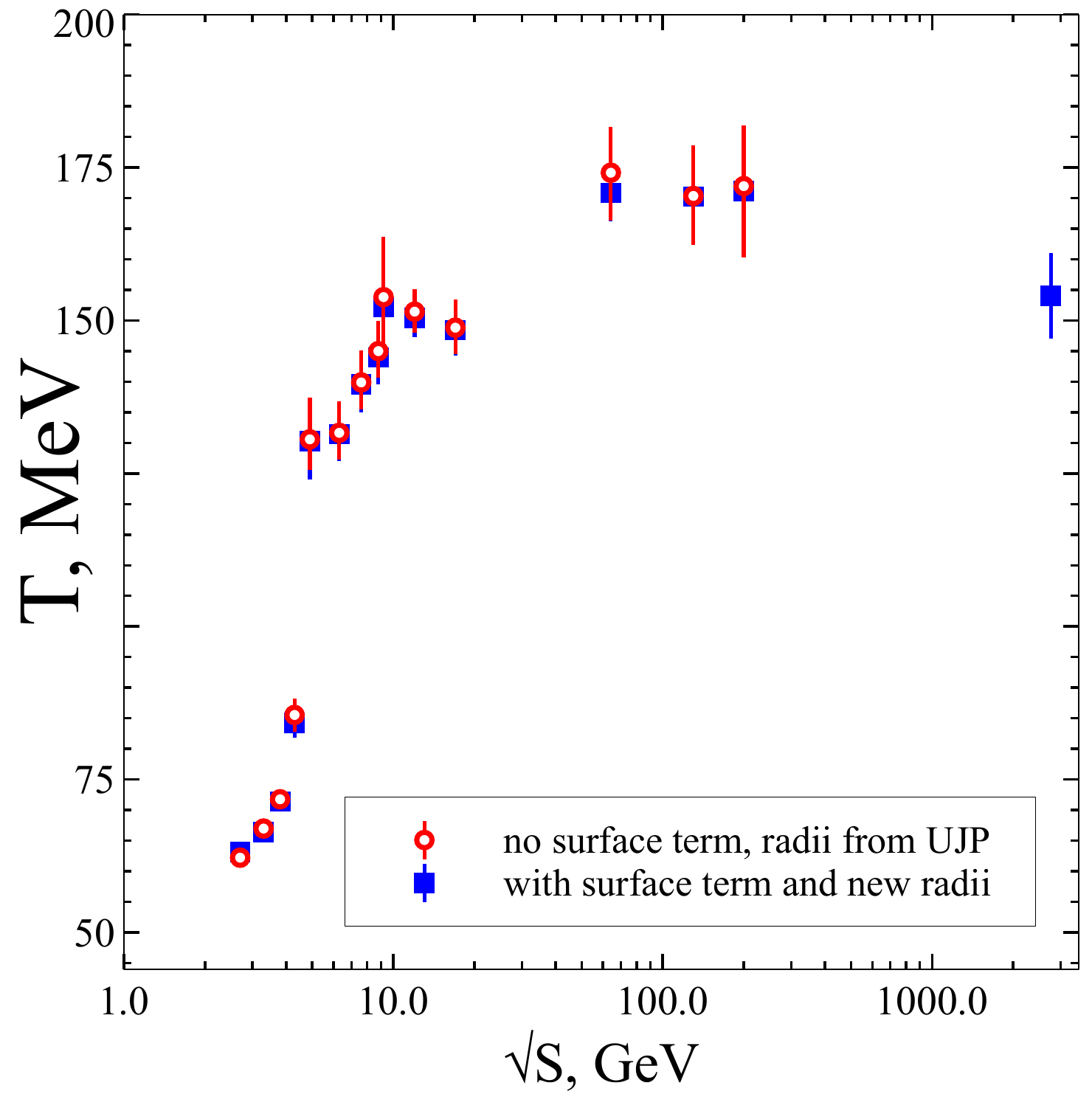}
  }
 \caption{Chemical freeze-out parameters of 
  the IST EoS  (circles) are compared to the  ones found within 
  the   HRGM with multicomponent hard-core repulsion (squares) \cite{Veta14}. 
Left  and right panels show, respectively,  the center of mass collision energy dependence of  the baryonic chemical potential 
and temperature.
 }
  \label{Fig4}
\end{figure}

\begin{figure}[htbp]
\centerline{
\includegraphics[width=77mm]{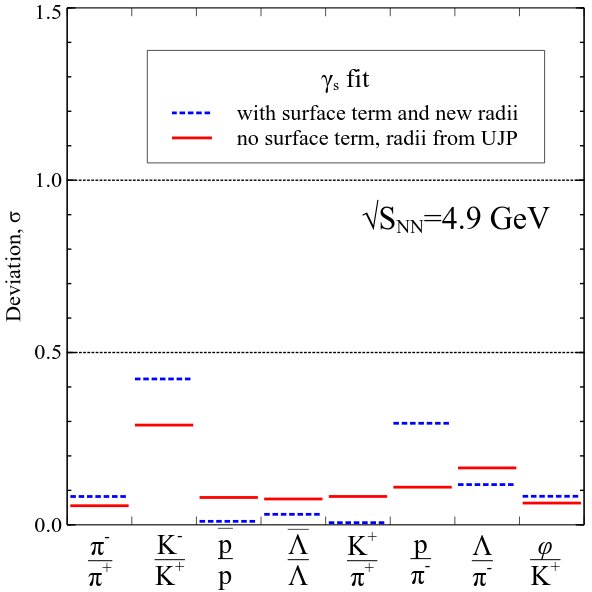}
  \hspace*{0.22cm}
\includegraphics[width=77mm]{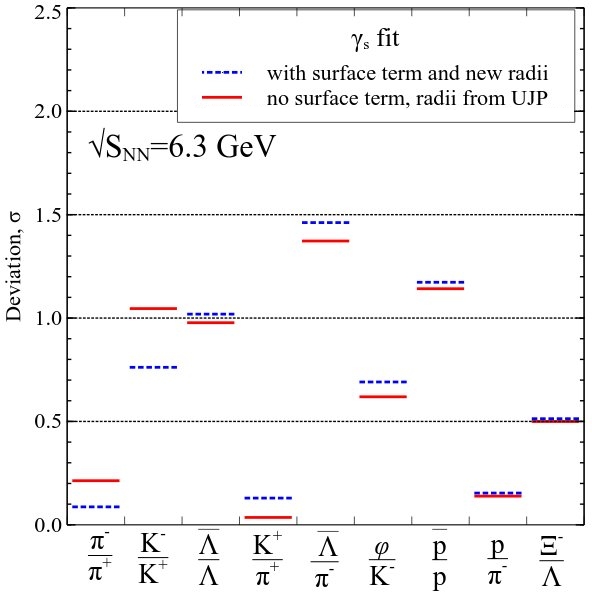}
  }
  \vspace*{11mm}
  \centerline{
\includegraphics[width=77mm]{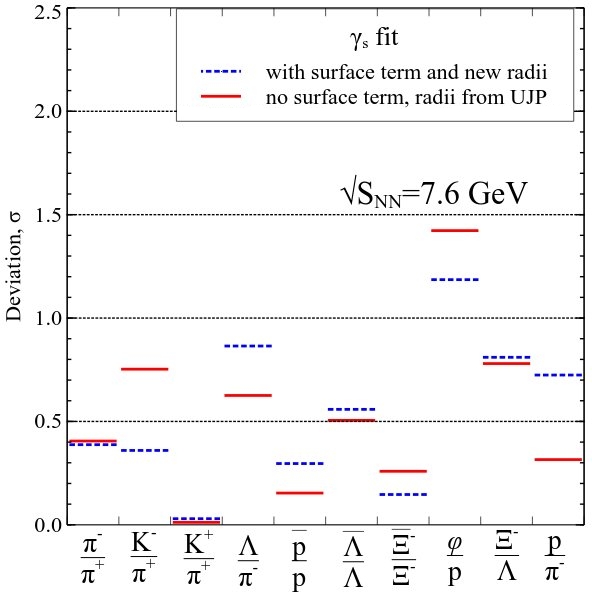}
  \hspace*{0.22cm}
\includegraphics[width=77mm]{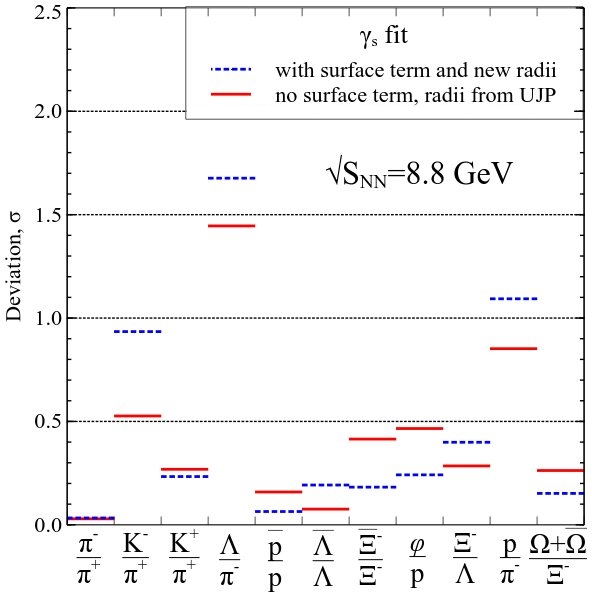}
  }
 \caption{Deviations of theoretically predicted hadronic yield ratios from experimental values in units of
 experimental error $\sigma$ are shown for the center of mass collision energies $\sqrt{s_{NN}} = 4.9, 6.3, 7.6, 8.8$ GeV. 
 Dashed lines correspond to the IST EoS fit, while the 
 solid lines correspond to the original HRGM  fit \cite{Veta14}. }
\label{Fig5}
\end{figure}

\begin{figure}[htbp]
\centerline{
\includegraphics[width=77mm]{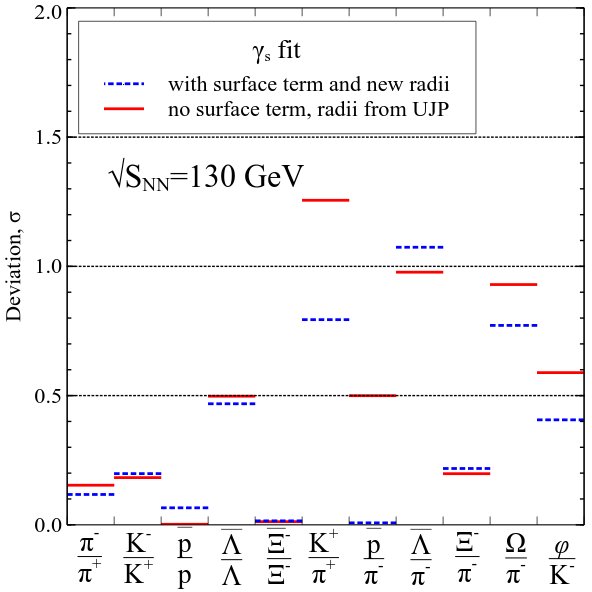}
  \hspace*{0.22cm}
\includegraphics[width=77mm]{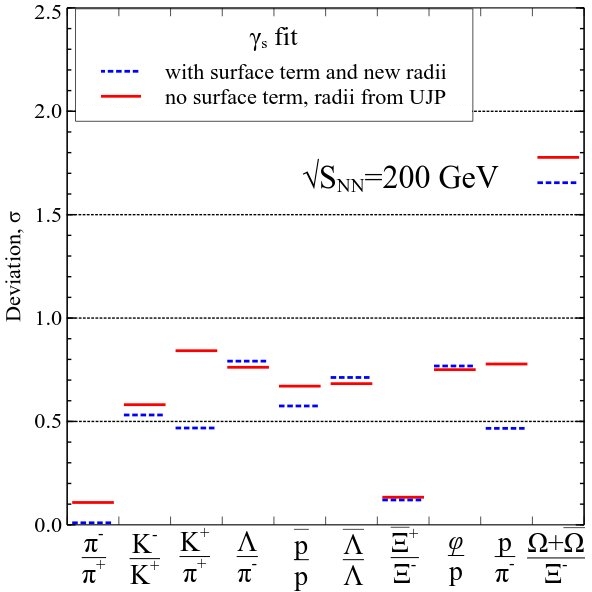}
  }
 \caption{ Same as in Fig. \ref{Fig5}, but for the center of mass collision energies
 $\sqrt{s_{NN}} = 130$ GeV and  $\sqrt{s_{NN}} = 200$ GeV. 
 }
\label{Fig6}
\end{figure}

The contribution of the resonance decays is  taken into  account   as usual: the total 
density of hadron $X$  consists of the thermal part  $n^{th}_X$  and the decay ones:
\begin{eqnarray}\label{EqXXVIII}
n^{tot}_X = n^{th}_X+ n^{decay} = n^{th}_X + \sum_{Y} n^{th}_Y \, Br(Y \to X) \,,
\end{eqnarray}
where $Br(Y \to X)$ denotes the decay branching of  the Y-th hadron  into the hadron X. The masses, the  widths and the strong decay branchings of all hadrons  were  taken from the particle tables  used  by  the  thermodynamic code THERMUS \cite{THERMUS}.

\subsection{Results for AGS, SPS and RHIC energies}

The parameter  $\alpha =1.25$ is used to fit 111 independent hadron ratios measured at AGS, SPS and RHIC energies with the  IST EoS. In this fit the factor $\gamma_s$ and the chemical potentials  $\mu_B$ and $\mu_{I3}$  are used as the free parameters  and  we found that the best description of these data is reached for  the following 
hard-core radii of baryons $R_{b}$=0.365 fm, mesons $R_{m}$=0.42 fm, pions $R_{\pi}$=0.15 fm, kaons
$R_{K}$=0.395 fm and $\Lambda$-hyperons $R_{\Lambda}$=0.085 fm. 
These values of the hard-core radii generate $\chi_1^2/dof=57.099/55 \simeq 1.038$. 

Compared to the values found by the HRGM \cite{Veta14},
i.e. the  hard-core radii of baryons  $R_{b}$=0.355 fm, mesons $R_{m}$=0.4 fm, pions $R_{\pi}$=0.1 fm, kaons
$R_{K}$=0.395 fm and $\Lambda$-hyperons $R_{\Lambda}$=0.11 fm, the hard-core radii of the IST EoS $R_{b}$, $R_{m}$ and $R_{K}$ are practically unchanged,  while the pionic hard-core radius is increased by 50\% and  the hard-core radius of  
$\Lambda$-hyperons is  diminished by 20\%.  From Fig.  \ref{Fig4} one can see that, despite the different hard-core radii of pions and   $\Lambda$-hyperons,  the collision energy dependence of the baryonic chemical potential and temperature at CFO are unchanged 
compared to the HRGM \cite{Veta14}.   The sudden jump of the CFO temperature observed between the collision energies  $\sqrt{s_{NN}} = 4.3$ GeV and  $\sqrt{s_{NN}} = 4.9$ GeV also remains unchanged.  This is an important finding since  such an irregularity, analyzed for the first time in \cite{Bugaev:2014}, led to a discovery of  possible  signals of the mixed phase formation in the central nuclear collisions \cite{Bugaev:2014,Bugaev:2015}.

Some typical results of the IST EoS fit are compared with the ones of HRGM in Figs.   \ref{Fig5} and  \ref{Fig6}.
As one can see from these figures at the  collision energies   $\sqrt{s_{NN}} = 4.9$ GeV,  $\sqrt{s_{NN}} = 6.3$ GeV
and  $\sqrt{s_{NN}} = 200$ GeV
the quality of the IST EoS  fit is almost the same as the one achieved with the  HRGM. At the  collision energies   $\sqrt{s_{NN}} = 7.6$ GeV  and  $\sqrt{s_{NN}} = 130$ GeV one can find  an improved description  of the  $\phi$-meson to proton ratio and the $K^+$-meson to $\pi^+$-meson ratio, respectively, while at  $\sqrt{s_{NN}} = 8.8$ GeV we found 
a slight worsening in the description of proton to $\pi^-$-meson ratio and  in the ratio $\Lambda/\pi^-$ (see Fig. \ref{Fig5}). 
The  fit results for other collision energies  obtained by the  HRGM and  by the IST EoS are hardly distinguishable from each other.

\begin{figure}[htbp]
\centerline{
\includegraphics[height=7 cm]{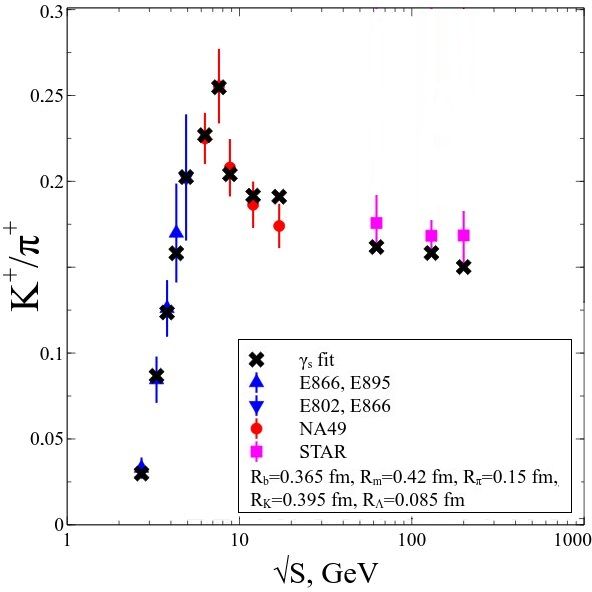}
\hspace*{1.cm}
\includegraphics[height=7.cm]{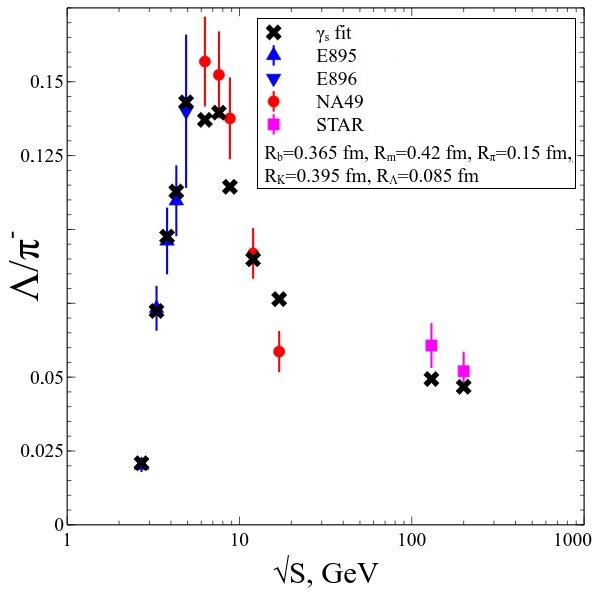}   
}
\centerline{
\includegraphics[height=7 cm]{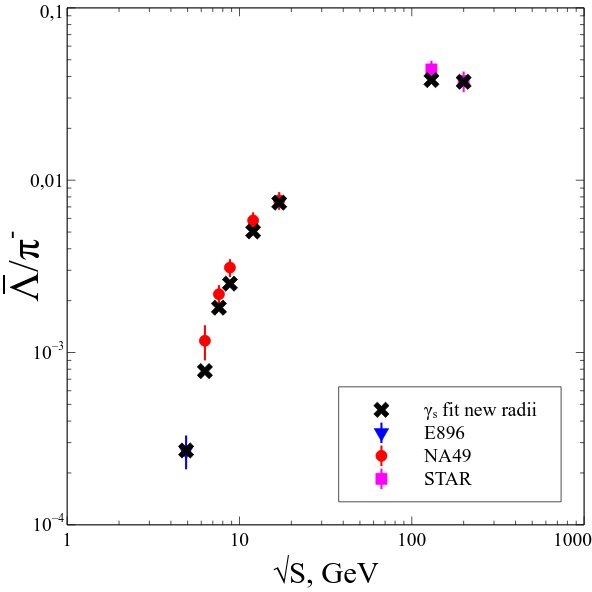}   
\hspace*{1.cm}
\includegraphics[height=7.cm]{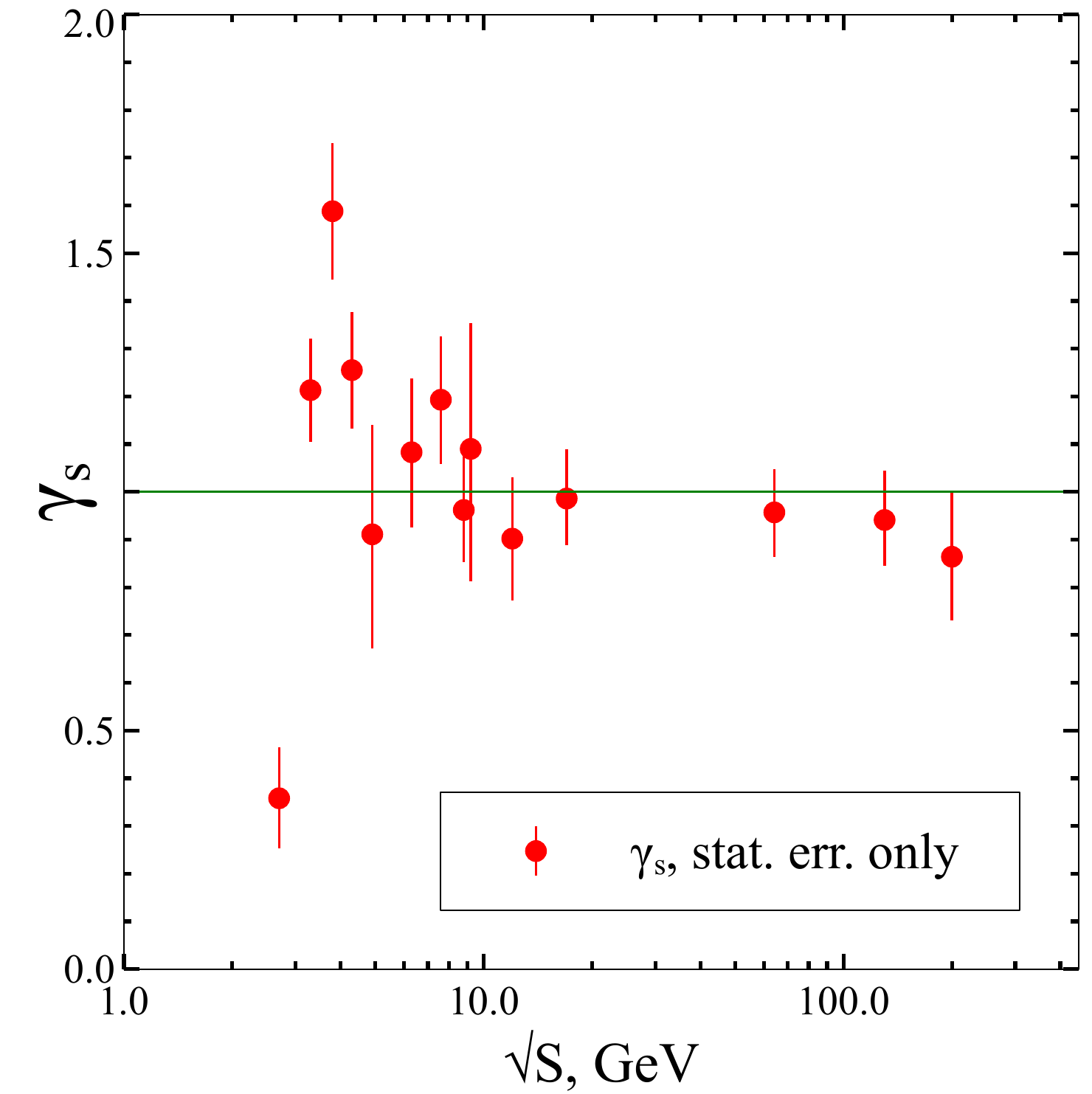}	
}	
 \caption{The fit results obtained by the IST EoS. {\bf Upper left panel:} $\sqrt{s_{NN}}$  dependence  of  $K^+/\pi^+$. {\bf Upper right panel:} $\sqrt{s_{NN}}$  dependence  of $\Lambda/\pi^{-}$. {\bf Lower left panel:}  $\sqrt{s_{NN}}$  dependence of  $\bar{\Lambda}/\pi^{-}$. {\bf Lower right panel:}  $\sqrt{s_{NN}}$  dependence  of the factor  $\gamma_{s}$.  
}
  \label{Fig7}
\end{figure}

We would like to mention  that the  IST EoS provides an improvement of     the $K^+/\pi^+$ description (the Strangeness Horn) from $\chi^2/dof \simeq 3.92/14$ in  \cite{Veta14} to $\chi^2/dof \simeq 3.29/14$ here, while $\sqrt{s_{NN}}$ dependences of 
$\Lambda/\pi^-$ and $\bar \Lambda/\pi^-$  ratios are reproduced here with $\chi^2/dof \simeq 11.62/12 $ and 
$\chi^2/dof \simeq 8.89/8$, respectively. Compared to the fit  qualities  $\chi^2/dof \simeq 10.22/12 $   for $\Lambda/\pi^-$ and 
$\chi^2/dof \simeq 6.49/8$ for $\bar \Lambda/\pi^-$  obtained in  \cite{Veta14},  
 the present results are slightly worse, but still they are rather good.  The collision energy dependence of these ratios is shown in  Fig. \ref{Fig7}. 
 
The other important finding is that  the collision energy dependence of the factor $\gamma_s$  for the IST EoS  is practically the same as for the HRGM of Ref. \cite{Veta14}. Thus,  the factor $\gamma_s$  demonstrates a low sensitivity
to the IST EoS, which means that the present model  confirms an existence of  a strangeness enhancement at low collision energies, namely the peak of the  factor $\gamma_s$  is found  at  $\sqrt{s_{NN}} = 3.8$ GeV as one can see from Fig. \ref{Fig7}.
Besides,  this figure shows that for $\sqrt{s_{NN}} \ge  4.9$ GeV there is chemical equilibrium of strange charge, since
$\gamma_s \simeq 1$ within the error bars. One possible explanation of  such a behavior is an appearance of quark gluon bags with the Hagedorn (exponential) mass spectrum \cite{Hagedorn}. Since such a mass spectrum acts as a perfect thermostat and 
a perfect chemical reservoir  \cite{Thermostat1} one may expect that the hadrons appearing   at 
the moment of quark gluon bag  hadronization will be born in a full thermal and chemical equilibrium \cite{Thermostat1,Hstate1,Hstate2,Hstate3}.
We would like to stress that this conclusion is in line with the recent finding that the mixed quark-gluon-hadron phase
is created  at the collision energy range between 4.3 GeV $ < \sqrt{s_{NN}} \le  4.9$ GeV \cite{Bugaev:2014,Bugaev:2015,Bugaev:2016a}.  Hence,  the observed change of the collision energy dependence  regime of  $\gamma_s$  at  
$\sqrt{s_{NN}} \simeq  4.9$ GeV maybe another evidence for the onset of deconfinement.

\subsection{Results for ALICE energy}

\begin{figure}
\centerline{
\includegraphics[width=\textwidth]{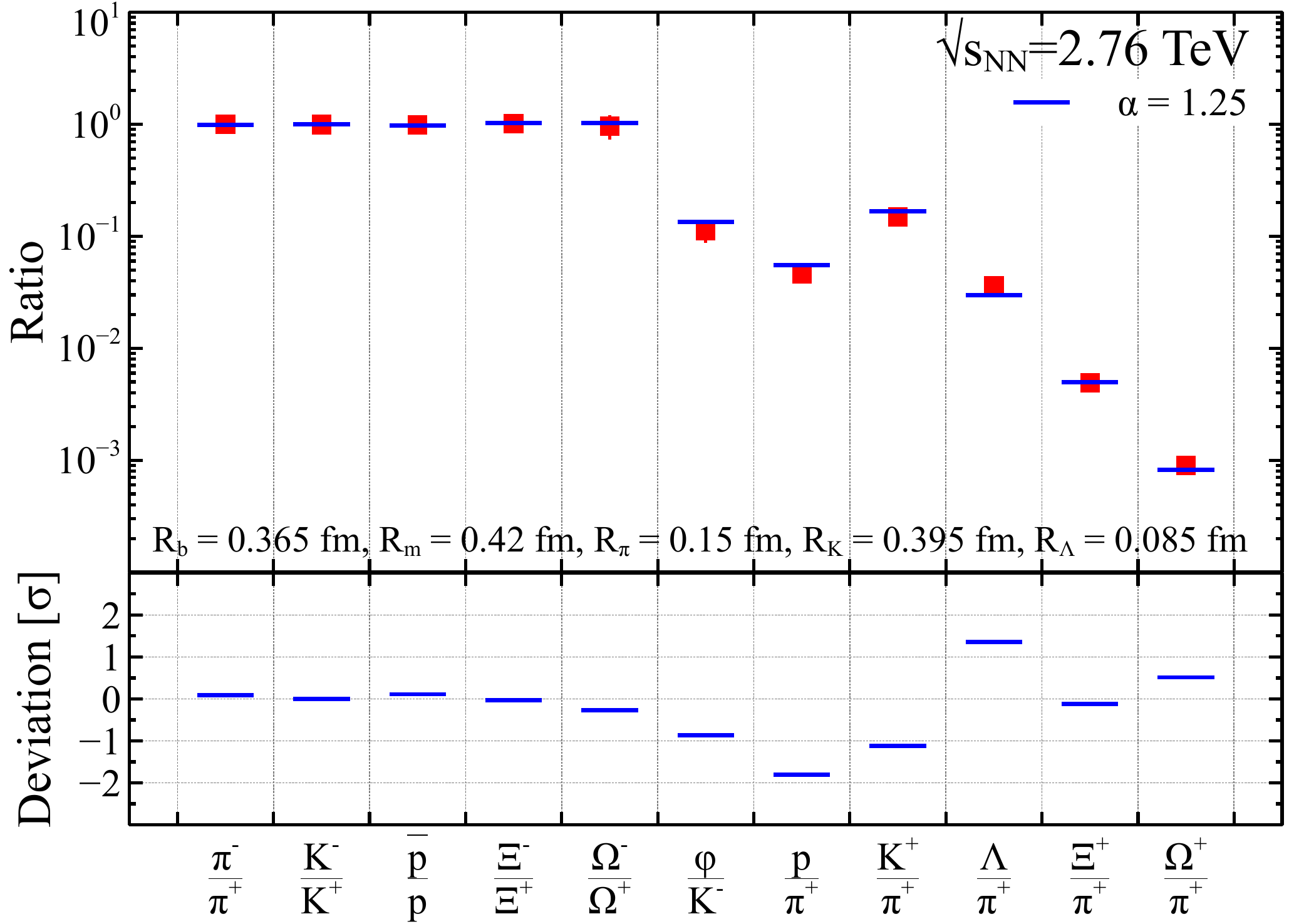}
}
 \caption{The ratios from Table 1 which were fitted by the IST EoS with the new radii found in this work. 
 The obtained  CFO temperature is  $T_{CFO} \simeq 152 \pm 7$ MeV. 
 The  quality  of the fit is  $\chi^2_2/dof \simeq 8.04/5 \simeq 1.61$. The upper panel shows the fit of the ratios, while the lower panel shows the deviation between data and theory in the units of  error.}
\label{Fig8}
\end{figure}

To fit  the ALICE data   \cite{Abelev:2013vea,Abelev:2013zaa,Abelev:2013xaa,Knospe:2013tda,Adam:2015vda,Donigus:2015bsa,Adam:2015yta} we use a different strategy. The reason is that the fit quality is not sensitive to the values of the hard-core radii. In fact, even the HRGM with the  point-like particles provides a reasonable fit quality \cite{Bugaev:2016,Chaterjee15}.
Therefore, in order to avoid the unnecessary  waste of  CPU time we adopted the new radii found in this work from fitting the 
AGS, SPS and RHIC data, then, similarly to \cite{Stachel:2013zma},  we set all values of chemical potentials  to zero, but the factor  $\gamma_s$  is fixed as $\gamma_s=1$. Thus, for the ALICE data we come up with 11 independent ratios (see Table 1)
and with  a single fitting parameter, namely the CFO temperature
which is found $T_{CFO} \simeq 152 \pm 7$ MeV.  
 Within the error bars this result is in agreement with the similar fits \cite{Stachel:2013zma,Chaterjee15}.  
 The achieved description of the ALICE data is shown in Fig. \ref{Fig8}.  The fit quality $\chi^2_2/dof \simeq 8.04/5 \simeq 1.61$  of the ALICE data is slightly worse than the one found for the combined fit of the AGS, SPS and RHIC data. From  Fig. \ref{Fig8} 
 one can see that the main part of  $\chi^2_2$ is generated by only two ratios, i.e. ${p}/{\pi^+}$ and  ${\Lambda}/{\pi^+}$.
 Therefore, the combined quality of the AGS, SPS, RHIC and ALICE data description achieved in the present work  is  $\chi^2_{tot}/dof \simeq 65.1/60 \simeq 1.08$.

 Although the  found CFO temperature for the ALICE data is rather low, we note that a priori it was not clear what   maximal value  for $T_{CFO}$ has to be chosen. For example, the authors of Ref. \cite{Vovch15}  claimed that they found the second minimum  of  $\chi^2/dof$  for the ALICE data which is located at the temperature about $274$ MeV. 
 Of course, it is hard to believe  that at so high temperature the hadrons  may exist and that at so  huge particle  densities the inelastic reactions  are frozen, but  in order not to miss  the $\chi^2/dof$ minimum at  high temperatures  we performed  its minimization  for $T_{CFO} < 600$ MeV. 
 
\begin{figure}[htb]
\centerline{
\includegraphics[width=77mm,height=68mm]{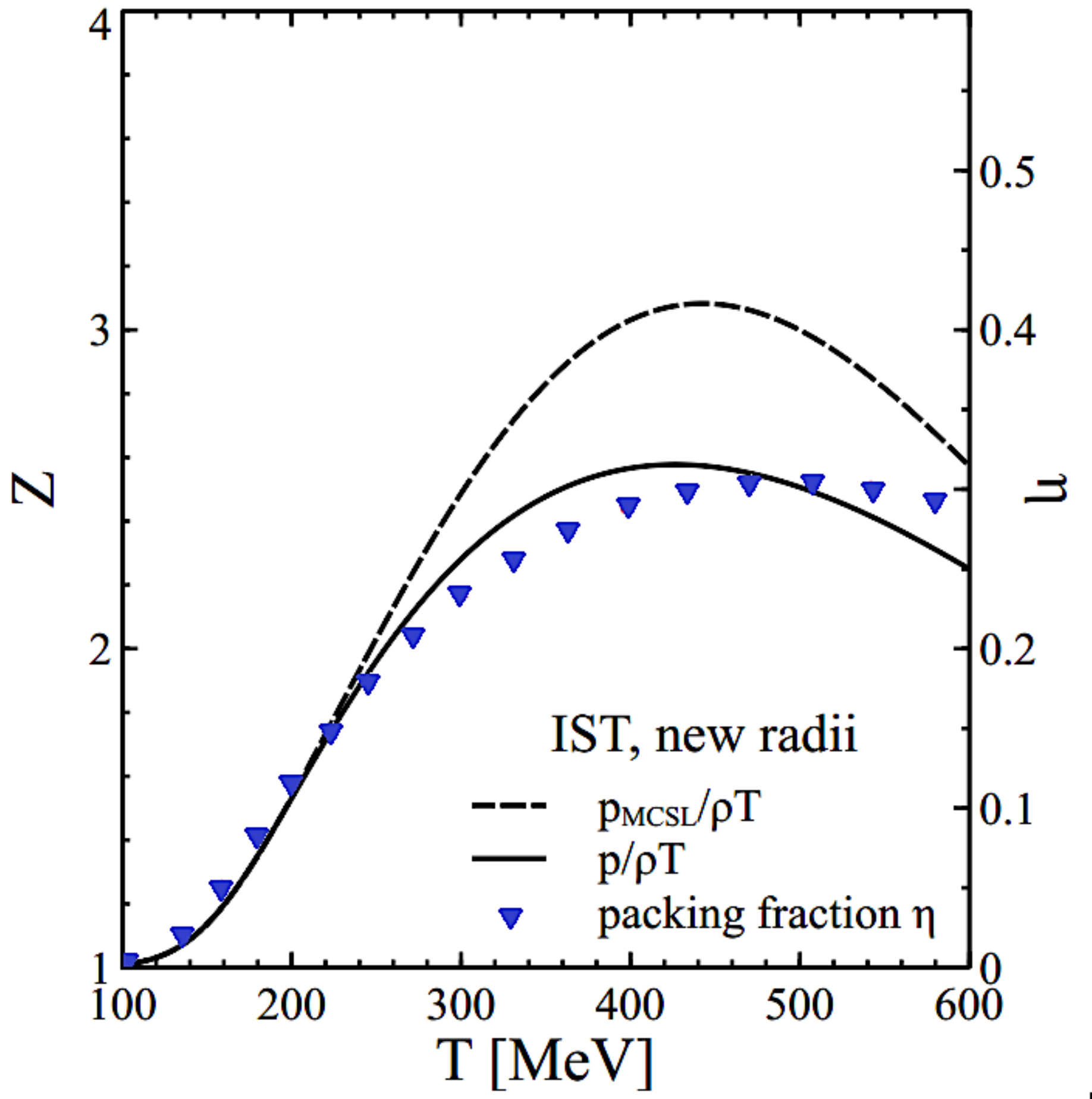}
  \hspace*{0.22cm}
\includegraphics[width=77mm,height=69mm]{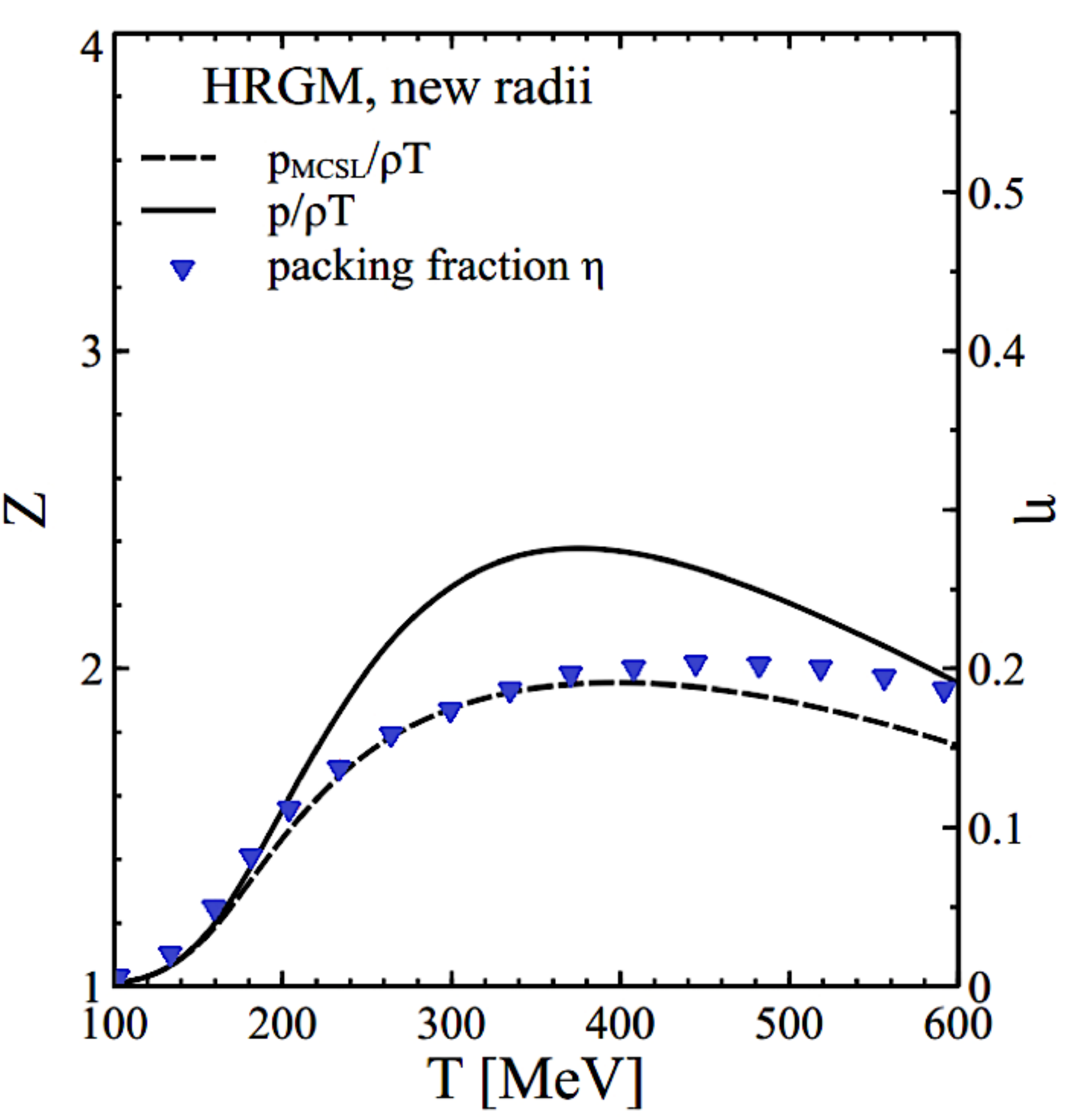}
  }
 \caption{{\bf Left panel:}  Comparison of the thermal compressibility $Z$ as the function of  CFO temperature $T$ is made for the IST EoS (solid curve)  and for the MCSL EoS (dashed  curve) found for the same particle densities for the hard-core radii of this work.  {\bf Right panel:}
A similar comparison between the  HRGM results obtained for the same set of hard-core radii   (solid curve)  and the MCSL EoS (dashed curve) is  shown.  }
\label{Fig9}
\end{figure}

 Although one can formally employ the IST EoS at any temperature, first we would like to determine the temperature range of its applicability. For this purpose
 we employ the multicomponent version  of the Carnahan-Starling EoS known as  the MCSL EoS \cite{CSmultic}.
 Such an EoS is   well known  in the theory of simple liquids \cite{SimpleLiquids1,SimpleLiquids2}.
 Similarly to its one-component counterpart  \cite{CSeos} the MCSL EoS rather accurately reproduces the pressure of hard spheres  until the packing fraction values  $\eta \le 0.35-0.4$ \cite{CSmultic,SimpleLiquids2}. As usual,  the packing fraction of the $N$-component mixture  $\eta \equiv  \sum\limits_{k=1}^N \frac{4}{3} \pi R_k^3 \rho_k$ is defined
 via the set of  hard-core radii $\{ R_k\}$ and the corresponding particle densities $\{ \rho_k\}$.  In terms of these notations  the MCSL  pressure  
 \cite{CSmultic} can be cast  as 
 \begin{eqnarray}\label{EqXXIX}
p^{CS} &=&  \frac{6\, T}{\pi} \left[  \frac{\xi_0}{1-\xi_3}  + \frac{3\, \xi_1 \xi_2}{(1-\xi_3)^2} + 
\frac{3\,  \xi_2^3}{(1-\xi_3)^3}  -  \frac{ \xi_3  \xi_2^3}{(1-\xi_3)^3} \right] \,,\\
\label{EqXXX}
\xi_n  &=&   \frac{\pi}{6}  \sum\limits_{k=1}^N \rho_k \left[ 2\, R_k \right]^n \,.
\end{eqnarray}
Using the system (\ref{EqXXIX}), (\ref{EqXXX}) we can find out the applicability bounds of  the IST EoS at high temperatures
by comparing the IST EoS  pressure (\ref{EqI}) with the MCSL pressure (\ref{EqXXIX}) which we calculate  for the  same 
set of particle densities  $\{\rho_k\}$ given by Eq. (\ref{EqXXI}). The results for the compressibility $Z = p/(\rho\, T)$ are given in Fig. \ref{Fig9}. Here  the total  pressure of the system is $p$, while the total particle density is $\rho =  \sum\limits_{k=1}^N \rho_k$.

From the left panel of  Fig. \ref{Fig9} one can see that  for $T \le  275$ MeV the IST EoS   obeys the condition applicability $\eta < 0.22$. Note also that at  $T \simeq 275$ MeV  the IST EoS
provides a 5\% deviation from the MSCL EoS at $T \simeq 275$ MeV, i.e. in the region where the second minimum of $\chi^2/dof$ is observed in the work   \cite{Vovch15}. 
But in contrast to Ref.  \cite{Vovch15},   we do not observe 
any additional minimum in our model up to $T = 600$ MeV.  A detailed analysis of  the ALICE data for different versions of the HRGM can be found in  \cite{Bugaev:2016}.

An entirely  different situation is for  the EVM. 
 In contrast to the IST EoS,  the EVM is stiffer than the MCSL EoS as one can see from the right panel of  Fig. \ref{Fig9}.
Also from this figure  one can see that the model  is not valid at high temperatures: 
the conventional  HRGM with multicomponent hard-core repulsion  is valid for packing fractions  $\eta \le 0.11$, i.e. for  $T < 200$ MeV.
From the right panel of  Fig. \ref{Fig9} one can find that this EoS 
provides a 5\% deviation from the MCSL EoS at  $T \simeq 215$ MeV. Therefore, we conclude that the HRGM  EoS
cannot be used  at higher temperatures  because  it  becomes  too stiff  even compared to the hard spheres  and, hence, it leads to the superluminal speed of sound.

\section{Conclusions}

In this work we discussed an entirely new EoS of hadronic matter and the results obtained by this model. 
The developed EoS is a physically transparent generalization of the multicomponent Van der Waals EoS 
which correctly accounts for  the third and fourth virial coefficients of the gas of hard spheres.  This is shown by the direct evaluation of the third and fourth virial coefficients for the one-component case, i.e. when all particles have the same hard-core radius.  
A single value of the  model parameter $\alpha=1.25$ allows us to simultaneously reproduce the third and fourth virial coefficients of 
the gas of hard spheres  with small deviations from their exact values. This very fact   shows that the developed model catches the correct physics and we explicitly  demonstrate that  the parameter $\alpha$ plays  the role of  a ``switcher"  between the 
eigenvolume and excluded volume regimes.
This conclusion is supported by a comparison of the compressibility factor $Z$ and speed of sound of the IST EoS and the ones 
found by the Carnahan-Starling EoS \cite{CSeos}  for the same particles  and same hard-core radii.  

Moreover, at vanishing
baryonic densities we compared
the compressibility factor $Z$   of the IST EoS and the one calculated for the multicomponent version of the Carnahan-Starling EoS \cite{CSmultic} known as the Mansoori-Carnahan-Starling-Leland EoS. This comparison shows that the conventional  multicomponent HRGM corresponds to the MCSL EoS at  the temperatures below $215$ MeV, while the IST EoS reproduces 
 the MCSL EoS  compressibility factor $Z$ with the relative error $5$\% up to the  temperature   $275$ MeV. 
 Furthermore, we found that in contrast to the conventional  HRGM   the IST EoS  is essentially softer than the Carnahan-Starling  and the MCSL EoS at high particle number densities. Thus,   the developed EoS breaks causality at 
 so high   densities that  in this region   the hadronic
 description should be replaced by the quark gluonic one.

Using the IST EoS we described the AGS, SPS, RHIC  and ALICE with rather high fit  quality $\chi^2_{tot}/dof  \simeq 1.08$.
Compared to the hard-core radii found within the HRGM \cite{Veta14} only the pion hard-core radius changed from $R_\pi = 0.1$ fm to $R_\pi = 0.15$ fm and the hard-core radius of $\Lambda$ (anti)hyperons decreased from $R_\Lambda = 0.11$ fm to $R_\Lambda= 0.085$ fm.  The other hard-core radii are almost the same as in   Ref. \cite{Veta14}. These values of hard-core radii allow us to describe all the hadronic yield ratios well, including the energy dependence of the  most problematic ones, i.e.  $K^+/\pi^+$,  $\bar{\Lambda}/\pi^{-}$ and  $\Lambda/\pi^{-}$ ratios. Despite the change of hard-core radii of pions and $\Lambda$ (anti)hyperons,
the present analysis confirmed that  the peak of  strangeness enhancement exists at $\sqrt{s_{NN}} = 3.8$ GeV and that  
the sudden jump of   the CFO temperature occurs  between $\sqrt{s_{NN}} = 4.3$ GeV and  $\sqrt{s_{NN}} = 4.9$ GeV.
 Furthermore, we argue that the appearance of chemical equilibrium of strangeness, i.e.  $\gamma_s \simeq 1$ within the error bars,  observed  at 
$\sqrt{s_{NN}} = 4.9$ GeV maybe a new signal  for  the onset of deconfinement.

It is necessary  to stress, that the great numerical  advantage of the IST EoS over the existing EVM is that  independently of the number of  different hard-core radii one has to solve just two equations.  Hence, we believe that such a property may open absolutely new possibilities in the future to extract the hard-core radii from the data with very high confidence, if the measurements at NICA and FAIR  will provide  much more precise  data for hadron multiplicities. 

\vspace*{2.2mm}

\noindent 
{\bf Acknowledgments.}
The authors  are   thankful to A. Andronic,  P. Braun-Munzinger, R. Emaus, M. Gazdzicki, D. R. Oliinychenko and  D. H. Rischke    for the fruitful discussions and for important  comments. K.A.B., V.V.S., A.I.I and G.M.Z. acknowledge the  partial support of the program ``Nuclear matter under extreme conditions''  launched by the Section of Nuclear Physics of National Academy of Sciences  of Ukraine. 
K.A.B. acknowledges the partial support by the ExtreMe Matter Institute EMMI, GSI Helmholtzzentrum f\"ur Schwerionenforschung, Darmstadt, Germany. 
The work of D.B. and A.I.I. was supported in part by the Polish National Centre (NCN) under contract number UMO-2011/02/A/ST2/00306. D.B. is grateful for support by the MEPhI Academic Excellence Project under contract
number 02.a03.21.0005.  C.G. acknowledges a support from HIC for FAIR.



\appendix

\section{Appendix}

In order to   heuristically derive the IST EoS, let us first 
remind  the formal steps of  obtaining the Van der Waals EoS in the grand canonical 
ensemble. For this purpose we  consider  first the one-component gas with the hard-core repulsion.  The pressure of such a gas  with the temperature $T$ and chemical potential $\mu$  in the Boltzmann  approximation is given by 
\begin{eqnarray}\label{EqIA}
p=T \, \phi (T) \exp\left[ \frac{\mu-a\, p}{T} \right] \,, \quad {\rm with} \quad \phi (T) = g \int \frac{d^3 \, 
 k}{(2\pi)^3}~\exp\left[  -\frac{ \sqrt{k^2 + m^2} }{T} \right] \,. 
\end{eqnarray}
Here  $a = \frac{2}{3} \pi (2\, R)^3$ denotes an excluded volume  per particle for the hard-core radius $R$ and  $\phi (T)$  is a thermal density 
of  hadron  having the mass $m$  and the degeneracy factor $g$. In general  case, which accounts for  the finite width of hadronic resonances  one has to use  
Eq.  (\ref{EqIV}) for a thermal density.
For  low particle number  densities Eq. (\ref{EqIA}) can be obtained from the virial expansion  at low densities 
\begin{eqnarray}
\label{EqIIA}
p \simeq  T \, 
\phi \, e^{\frac{\mu}{T}}\left(1-a\, \phi \, e^{\frac{\mu}{T}}\right) \,, 
\end{eqnarray}
in the following sequence of steps \cite{Bugaev:13NPA}. First, in the second term staying in the brackets in  Eq. (\ref{EqIIA})  one has to approximate the particle density as $\phi \, e^{\frac{\mu}{T}} \simeq \frac{p}{T} $, using the fact that at low densities such an approximation is a correct one; second, the obtained term is further approximated as  $1 - a\, \frac{p}{T} \simeq \exp\left[ - a\, \frac{p}{T}  \right] $. As a  result Eq. (\ref{EqIA}) is reproduced. The final step is to extrapolate Eq. (\ref{EqIA}) to high densities.

Let us  now apply the same steps  to the system of $N$-sorts  of  hadrons  with  the hard-core radii $R_k$, with $k = 1, 2, ..., N$. Then the virial expansion of the gas pressure  up to second order in particle density  can be written as
\begin{eqnarray}\label{EqIIIA}
p   =T \sum_{k=1}^{N} \phi_{k}\, e^{\frac{{\mu}_{k}}{T}}\left(1- \sum_{n=1}^{N}\, a_{kn }\,  \phi_{n} \, e^{\frac{{\mu}_{n}}{T}} \right) \,,
\end{eqnarray}
where  $\phi _k(T)$ (\ref{EqIV}) is  the thermal density of particles of the degeneracy $g_k$ and mass $m_k$,  $\mu_k$ denotes their  chemical potential,  while  
$a_{kn}$ is  the excluded volume of  particles  having the hard-core-radii $R_k$ and $R_n$
\begin{eqnarray}\label{EqIVA}
a_{kn}= \frac{2}{3} \pi \left(R_{k}+R_{n}\right)^{3}=\frac{2}{3}  \pi \left(R_{k}^{3}+3R_{k}^{2}R_{n}+3R_{k}R_{n}^{2}+R_{n}^{3} \right).
\end{eqnarray}
One can  repeat the same steps as above  with the only modification that
for the multicomponent system  each sort of particles, say $k$,  generates its own pressure $p_k$ which 
should replace  in Eq. (\ref{EqIIIA}) the particle density as $ \phi_{n} \, e^{\frac{{\mu}_{n}}{T}} \simeq \frac{p_n}{T}$. Then one obtains the system of  equations for partial pressures  
\begin{eqnarray}\label{EqVA}
p_k  =  T \,  \phi_{k}\, \exp \left[ \frac{{\mu}_{k}}{T} -  \sum_{n=1}^{N}\, a_{kn }\,  \frac{p_n}{T}  \right]  \,.
\end{eqnarray}
Such an EoS  is known as the Lorentz-Berthelot mixture  for which the total pressure is the sum of all partial ones $p = \sum_{k=1}^{N}\, p_k$.  

However, the procedure of  the Van der Waals extrapolation is not unique and one can use a different approach. 
It is necessary to stress  that an order of  mathematical operations is important now \cite{Bugaev:13NPA}.  If, in contrast to 
the steps above,  one  explicitly substitutes the excluded volume  (\ref{EqIV}) into  expression for pressure  (\ref{EqIIIA})  first  and   regroups the powers of  radius $R_k$ of  a  
hadron of sort $k$, then one can  get a different expression
\begin{eqnarray}
\label{EqVIA}
p=T\sum_{k=1}^{N} \phi_{k}e^{\frac{{\mu}_{k}}{T}}\left[1-
\frac{4}{3}\pi R_{k}^{3}\cdot\sum_{n=1}^{N}\phi_{n}e^{\frac{{\mu}_{n}}{T}}-
2\pi R_{k}^{2}\cdot \sum_{n=1}^{N}R_{n} \phi_{n}e^{\frac{{\mu}_{n}}{T}}-
2\pi R_{k}\cdot\sum_{n=1}^{N}R_{n}^{2} \phi_{n}e^{\frac{{\mu}_{n}}{T}}\right] \,. 
\end{eqnarray}
Noting that the third and the fourth  terms on the right hand side  of  Eq.  (\ref{EqVIA}) are identical for low densities,
we can write  (\ref{EqVIA}) as
\begin{eqnarray}
\label{EqVIIA}
p=T\sum_{k=1}^{N} \phi_{k}e^{\frac{{\mu}_{k}}{T}}\left[1-
\frac{4}{3}\pi R_{k}^{3}\cdot\sum_{n=1}^{N}\phi_{n}e^{\frac{{\mu}_{n}}{T}}-
4\pi R_{k}^{2}\cdot \sum_{n=1}^{N}R_{n} \phi_{n}e^{\frac{{\mu}_{n}}{T}} \right] \,. 
\end{eqnarray}
Next we rewrite  the terms staying in the square brackets  in Eq. (\ref{EqVIIA}) via an exponential   
and  obtain 
\begin{eqnarray}\label{EqVIIIA}
p&=& T\sum_{k=1}^{N} \phi_{k}
\exp\left[ \frac{{\mu}_{k}}{T} -\frac{4}{3}\pi R_{k}^{3}\cdot\frac{p}{T}-4\pi R_{k}^{2}\cdot \frac{\Sigma}{T} \right] \,, 
\end{eqnarray}
where we made the same approximation for total pressure as above $T \sum_{n=1}^{N}\phi_{n}e^{\frac{{\mu}_{n}}{T}} \simeq p$. A similar approximation has to be made for the induced surface tension  coefficient $\Sigma$
in order to guarantee a consistency   with the  derivation above  and to have the   correct   values for  all second virial coefficients at low densities.  Hence  we assume that   the induced  surface free energy  coefficient  $\Sigma$ obeys  the following equation
\begin{eqnarray}
\label{EqIXA}
\Sigma & =  & T  \sum_{k=1}^N  R_k \, \phi_{k} \, \exp\left[ \frac{{\mu}_{k}}{T} -\frac{4}{3}\pi R_{k}^{3}\cdot\frac{p}{T}-4\pi R_{k}^{2}\cdot \alpha\, \frac{\Sigma}{T} \right] \,,
\end{eqnarray}
which is a complete analog of the equation (\ref{EqVIIIA}) for pressure. The only difference with  Eq. (\ref{EqVIIIA}) is the presence of   the constant $\alpha > 0$  which  is introduced due to the freedom of the Van der Waals extrapolation to high densities.  By construction the finite values of  $\alpha$ cannot affect the second virial coefficients, but
with it help  the present model is able to  account for higher order corrections compared  to the low density virial expansion. 
 
From the derivation above 
it is clear  that the induced surface tension coefficient $\Sigma > 0$ is generated by the hard-core repulsion and  accounts for its 
essential  part. As it  is argued in \cite{Bugaev:13NPA} the attractive interaction will lead to $\Sigma < 0$.



\begin{thebibliography}{99}

\bibitem{GenRevEOS}
N. K.  Glendenning, ``Compact Stars", Springer-Verlag, New York   (2000). 


\bibitem{Haensel:2007yy} 
  P.~Haensel, A.~Y.~Potekhin and D.~G.~Yakovlev,
  Astrophys.\ Space Sci.\ Libr.\  {\bf 326} (2007).

\bibitem{Mott:1974}
N.~F.~Mott, {\it Metal - Insulator Transitions}, Taylor \& Francis Ltd., New York (1974).  

\bibitem{CSeos}
N. F. Carnahan and  K. E. Starling, J. Chem. Phys. {\bf  51}, 635  (1969).


\bibitem{Ebeling:2008mg} 
  W.~Ebeling, D.~Blaschke, R.~Redmer, H.~Reinholz and G.~R\"opke,
  J.\ Phys.\ A {\bf 42}, 214033 (2009).

\bibitem{EbelingMottBook} 
  W.~Ebeling, D.~Blaschke, R.~Redmer, H.~Reinholz and G.~R\"opke,
  in: Metal-to-Nonmetal Transitions, Springer Series in Materials Science, vol. 132, 
  Springer, Berlin (2010), p. 37.


\bibitem{Lattimer:1991nc} 
  J.~M.~Lattimer and F.~D.~Swesty,
  Nucl.\ Phys.\ A {\bf 535}, 331 (1991).

\bibitem{Shen:1998gq} 
  H.~Shen, H.~Toki, K.~Oyamatsu and K.~Sumiyoshi,
  Nucl.\ Phys.\ A {\bf 637}, 435 (1998).

\bibitem{Typel:2009sy}
  S.~Typel, G.~R\"opke, T.~Kl\"ahn, D.~Blaschke and H.~H.~Wolter,
   Phys.\ Rev.\ C {\bf 81}, 015803 (2010).

\bibitem{Benic:2014jia}
  S.~Benic, D.~Blaschke, D.~E.~Alvarez-Castillo, T.~Fischer and S.~Typel,
  Astron.\ Astrophys.\  {\bf 577}, A40 (2015).


\bibitem{PBM06}
%
 A. Andronic, P. Braun-Munzinger and  J. Stachel, Nucl. Phys. A {\bf 772},  167 (2006) and references therein.
 
\bibitem{Stachel:2013zma}
  J.~Stachel, A.~Andronic, P.~Braun-Munzinger and K.~Redlich,
  J.\ Phys.\ Conf.\ Ser.\  {\bf 509}, 012019 (2014) and references therein.

\bibitem{Typel16}
%
S. Typel, Eur. Phys. J. A {\bf 52}, 16 (2016). 
 
  \bibitem{Horn}
 %
K. A. Bugaev, D. R. Oliinychenko,   A. S. Sorin and G. M. Zinovjev, 
Eur. Phys. J. A {\bf 49}, 30  (2013). 
 
\bibitem{KABOliinychenko:12}
%
D. R. Oliinychenko, K. A. Bugaev and  A. S. Sorin, 
Ukr. J. Phys.  {\bf 58},  211 (2013). 
 
 \bibitem{SFO}
 K.~A.~Bugaev et al., 
 Europhys. Lett. {\bf 104},  22002  (2013).
 
  \bibitem{Veta14}
 V. V. Sagun, Ukr. J. Phys. {\bf 59}, 755 (2014). 

 
\bibitem{Bugaev:2014}
  K.~A.~Bugaev et al., 
 Phys. Part. Nucl. Lett. {\bf 12},   351 (2015). 
 
\bibitem{Bugaev:2015}
  K.~A.~Bugaev et al., 
 Eur. Phys. J. A  {\bf 52}, 175 (2016). 
  
 \bibitem{Bugaev:2016}
 %
K. A. Bugaev  et al.,   arXiv:1611.07349v2  [nucl-th].

 \bibitem{Bugaev:2016ujp}
 %
 K. A. Bugaev et al., 
Ukr. J. Phys.  {\bf 61},  659 (2016).
  
 \bibitem{Bugaev:2016a}
 %
 K. A. Bugaev et al., 
 Eur. Phys. J. A  {\bf 52}, 227 (2016). 
 
\bibitem{Vovch15}
%
V. Vovchenko and  H. St{\"o}cker, arXive:1512.08046v2 [hep-ph]. 
  
\bibitem{SSpeed}
%
L. M. Satarov, K. A. Bugaev and I. N. Mishustin, 
Phys. Rev. C {\bf 91}, 055203 (2015).

\bibitem{QGBST3crit}
K. A. Bugaev, 
{Phys. Rev.} {\bf C 76},  014903  (2007). 

\bibitem{QGBSTcrit}
%
K. A. Bugaev, V. K. Petrov and G. M. Zinovjev,
Phys.  Atom. Nucl.  {\bf 76},  341  (2013).


\bibitem{Carsten07}
I. Zakout, C. Greiner, and J. Schaffner-Bielich,  Nucl. Phys. A
{\bf  781}, 150 (2007).

\bibitem{Carsten08}
 I. Zakout and C. Greiner, Phys. Rev. C  {\bf  78}, 034916 (2008).
 
 \bibitem{Ferroni}
L. Ferroni and V. Koch, Phys. Rev. C {\bf  79}, 034905 (2009).
 
 \bibitem{Carsten10}
I. Zakout and C. Greiner, arXiv:1002.3119.
 

\bibitem{IvanytskyiNPA}
A. I. Ivanytskyi, Nucl. Phys. A {\bf 880}, 12 (2012).


\bibitem{IvanytskyiPRE}
%
A. I. Ivanytskyi, K. A. Bugaev, A. S. Sorin and G. M. Zinovjev, Phys. Rev. E {\bf 86}, 061107  (2012).
 
\bibitem{UJP2012}
%
A. I. Ivanytskyi and K. A. Bugaev,
Ukr. J. Phys.  {\bf 57},  964 (2012).

  
\bibitem{Bugaev:13NPA} 
V. V. Sagun, K. A. Bugaev, A. I. Ivanytskyi, I.N. Mishustin,
Nucl. Phys. A {\bf 924}, 24 (2014).

\bibitem{Mekjian}
S. Das Gupta and A. Z. Mekjian, 
Phys. Rev. C {\bf 57}, 1361 (1998).  


\bibitem{David:16A}
 %
D. Blaschke, A. Dubinin and L. Turko, arXiv:1611.09845 [hep-ph].

 \bibitem{David:16B}
 %
D.~Blaschke, A.~Dubinin and L.~Turko,
  arXiv:1612.09556 [hep-ph].

 \bibitem{Phi-approach}
%
G. Baym, Phys. Rev. {\bf 127}, 1391 (1962).

 \bibitem{Phi-approach2}
%
B. Vanderheyden and  G. Baym, J. Stat. Phys.  {\bf  93}, 843 (1998).


 \bibitem{Rafelski}
 %
 J. Rafelski, Phys. Lett. B 62, 333 (1991).

 
\bibitem{BookLiquids}
%
J. P. Hansen and I. R. McDonald,  {\it ``Theory of simple fluids"}, Academic Press, Amsterdam, 2006.

 \bibitem{Abelev:2013vea}
  B.~Abelev {\it et al.} [ALICE Collaboration],
  Phys.\ Rev.\ C {\bf 88}, 044910 (2013).

\bibitem{Abelev:2013zaa}
  B.~B.~Abelev {\it et al.} [ALICE Collaboration],
  Phys.\ Lett.\ B {\bf 728}, 216 (2014);
  Erratum: [Phys.\ Lett.\ B {\bf 734}, 409 (2014)]

\bibitem{Abelev:2013xaa}
  B.~B.~Abelev {\it et al.} [ALICE Collaboration],
  Phys.\ Rev.\ Lett.\  {\bf 111}, 222301 (2013).

\bibitem{Knospe:2013tda}
  A.~G.~Knospe [ALICE Collaboration],
  J.\ Phys.\ Conf.\ Ser.\  {\bf 509}, 012087 (2014).

\bibitem{Adam:2015vda}
  J.~Adam {\it et al.} [ALICE Collaboration],
  Phys.\ Rev.\ C {\bf 93}, 024917 (2016).

\bibitem{Donigus:2015bsa}
  B.~D\"onigus [ALICE Collaboration],
  EPJ Web Conf.\  {\bf 97}, 00013 (2015).

\bibitem{Adam:2015yta}
  J.~Adam {\it et al.} [ALICE Collaboration],
  Phys.\ Lett.\ B {\bf 754}, 360 (2016).

\bibitem{THERMUS}
%
S. Wheaton, J. Cleymans and M. Hauer, Comput. Phys. Commun. {\bf 180}, 84 (2009).

\bibitem{Hagedorn}
%
R. Hagedorn, Nuovo Cim. Suppl.  {\bf 3}, 147 (1965).

\bibitem{Thermostat1}
%
L. G. Moretto,  K. A. Bugaev, J. B. Elliott and L. Phair,
 Europhys. Lett.  {\bf 76},  402 (2006).
%


\bibitem{Hstate1}
%
  M.~Beitel, K.~Gallmeister and C.~Greiner,
  Phys.\ Rev.\ C {\bf 90}, 045203 (2014).

\bibitem{Hstate2}
%
M.~Beitel, K.~Gallmeister and C.~Greiner,
  J.\ Phys.\ Conf.\ Ser.\  {\bf 668},  012057 (2016).

\bibitem{Hstate3}
%
M.~Beitel, C.~Greiner and H.~Stoecker,
  Phys.\ Rev.\ C {\bf 94},  021902 (2016).


\bibitem{Chaterjee15}
%
S. Chatterjee et al.,  Adv. High Energy Phys. {\bf 2015}, 349013 (2015).

\bibitem{CSmultic}
 %
 G. A. Mansoori, N. F. Carnahan, K. E. Starling and T. W. Leland, Jr., J. Chem. Phys. {\bf 54}, 1523 (1971).
 
 
 \bibitem{SimpleLiquids1}
%
J. P. Hansen and I. R. McDonald, {\it  ``Theory of  simple liquids"}, Academic, London (2006). 

 \bibitem{SimpleLiquids2}
%
J. J. Salacuse and G. Stell, 
J. Chem. Phys.  {\bf 77}, 3714 (1982).

%

  
\end{thebibliography}
\end{document}